# Quantum phase transitions between a class of symmetry protected topological states

Lokman Tsui [a], Hong-Chen Jiang [c], Yuan-Ming Lu [d], Dung-Hai Lee [a,b,*]

[a] *Department of Physics, University of California, Berkeley, CA 94720, USA*
[b] *Materials Science Division, Lawrence Berkeley National Laboratories, Berkeley, CA 94720, USA*
[c] *Stanford Institute for Materials and Energy Sciences, SLAC National Accelerator Laboratory, 2575 Sand Hill Road, Menlo Park, CA 94025, USA*
[d] *Department of Physics, The Ohio State University, Columbus, OH 43212, USA*

**Abstract**

The subject of this paper is the phase transition between symmetry protected topological states (SPTs). We consider spatial dimension $d$ and symmetry group $G$ so that the cohomology group, $H^{d+1}(G, U(1))$, contains at least one $Z_{2n}$ or $Z$ factor. We show that the phase transition between the trivial SPT and the root states that generate the $Z_{2n}$ or $Z$ groups can be induced on the boundary of a $(d+1)$-dimensional $G \times Z_2^T$-symmetric SPT by a $Z_2^T$ symmetry breaking field. Moreover we show these boundary phase transitions can be "transplanted" to $d$ dimensions and realized in lattice models as a function of a tuning parameter. The price one pays is for the critical value of the tuning parameter there is an extra non-local (duality-like) symmetry. In the case where the phase transition is continuous, our theory predicts the presence of unusual (sometimes fractionalized) excitations corresponding to delocalized boundary excitations of the non-trivial SPT on one side of the transition. This theory also predicts other phase transition scenarios including first order transition and transition via an intermediate symmetry breaking phase.

* Corresponding author at: Department of Physics, University of California, Berkeley, CA 94720, USA.
*E-mail addresses:* lokman@berkeley.edu (L. Tsui), hcjiang@slac.stanford.edu (H.-C. Jiang), lu.1435@osu.edu (Y.-M. Lu), dunghai@berkeley.edu (D.-H. Lee).

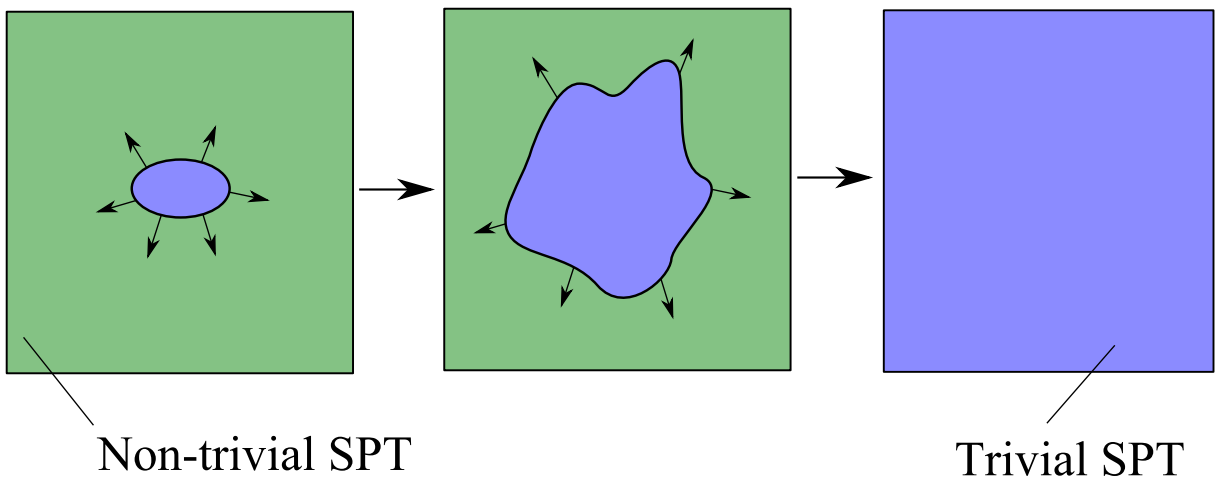

Fig. 1. (Color online.) A caricature showing the necessity of gapless excitations on the boundary of a non-trivial SPT. The blue and green regions represent a trivial and a non-trivial SPT respectively. If the interface between a trivial and non-trivial SPT were gapped, then a small island of trivial SPT may be grown inside a non-trivial SPT, and gradually expand to occupy the entire system without closing the energy gap hence adiabatically connecting a trivial and non-trivial SPT.

## 1. Introduction

Symmetry protected topological(SPT) phases are the non-degenerate ground states of local lattice Hamiltonians each respecting the same global symmetry group $G$. These ground states remain invariant under $G$ and are separated from their respective excited states by an energy gap. If two Hamiltonians can be made equal by adding or removing symmetry preserving local terms while preserving the excitation gap, they are viewed as equivalent. Correspondingly ground states of equivalent Hamiltonians are viewed as the same phase. Under the equivalence relation defined above, for fixed spatial dimension $d$ and symmetry group $G$, it is proposed [1] that the SPT phases form a 1-1 correspondence with an abelian group – the cohomology group $H^{d+1}(G, U(1))$. The elements are the different SPT phases and the group operation between two SPT phases corresponds to physically stacking the representative states of the two SPT phases in question on top of each other. The identity element is the trivial SPT phase, namely the phase whose equivalence class contains the direct product state. In the rest of the paper we will focus on bosonic problems.

The hallmark of non-trivial SPTs is the presence of gapless boundary excitations. The fact that a non-trivial SPT must have a gapless boundary can be understood as follows. An SPT with boundary can be alternatively viewed as the same SPT interfaces with vacuum, i.e., a trivial SPT. If there were no gapless excitations at the interface we can gradually expand an island of trivial SPT embedded in the non-trivial one until it occupies the entire system without closing the energy gap (see Fig. 1). Since such expansion, or more precisely the local modification of the Hamiltonian which causes such expansion, does not have to break the symmetry, this contradicts the notion of trivial and non-trivial SPT being in two inequivalent classes. Put it simply, the gap closure at the boundary between two inequivalent SPTs can be viewed as a spatial coordinate tuned phase transition between the two SPTs [2].

At the present time SPTs and their classification are largely understood. However, recently some exceptional phases are discovered outside the cohomology classification [3]. In the paper we focus on a new frontier – the phases transition between the SPTs classified by the cohomology group [4,5]. More precisely we imagine the bulk Hamiltonian has a tuning parameter $\lambda$, and by changing $\lambda$ the ground state of $H(\lambda)$ goes from one SPT to another inequivalent SPT. Our purpose is to study the possible phase transition(s) occurring for intermediate values of $\lambda$. The main result of this paper is the theorem associated with the following theorem.

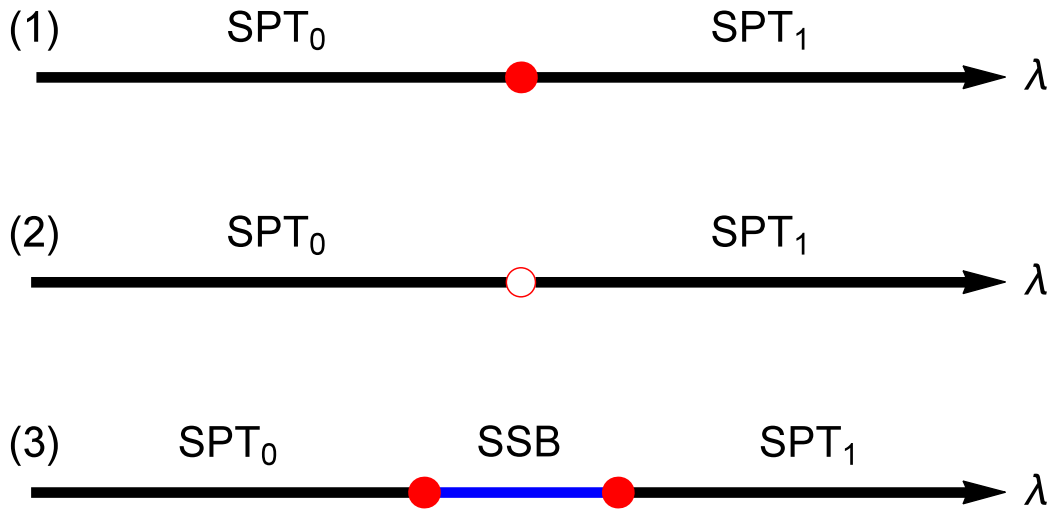

Fig. 2. (Color online.) Three possible scenarios for the phase transition between two different SPTs. Red dots represent continuous quantum critical points, red circle represents first order phase transition, and "SSB" stands for spontaneous symmetry breaking.

**Theorem.** *The three scenarios of phase transition (see Fig. 2) between a trivial $d$-dimensional $G$-symmetric SPT and a non-trivial SPT satisfying a special condition can be realized at the boundary a $(d+1)$-dimensional $G \times Z_2^T$ symmetric SPT under the influence of a* boundary $Z_2^T$ *symmetry breaking field. The condition the non-trivial $G$-symmetric SPT must satisfy is that it is not equivalent to the stacking of any two other identical $G$-symmetric SPTs. This condition will be referred to as the "non-double-stacking condition" (NDSC) in the rest of the paper. Any $G$ whose $H^{d+1}(G, U(1))$ contains a $Z_{2n}$ or $Z$ factor will have SPTs, e.g., that corresponds to the generator of $Z_{2n}$ or $Z$, satisfy this condition.*

Here the $Z_2^T$ transformation inverts the sign of a local Ising variable and performs a complex conjugation on the wavefunction. Because the Ising variable in question is not necessarily time reversal odd, this $Z_2^T$ is not the usual time reversal symmetry. This theorem allows us to construct explicit lattice models to describe the SPT phase transition. In particular these lattice models possess a non-local transformation (a "duality transformation") relating the trivial and non-trivial SPTs on the opposite sides of the transition. In the case of continuous phase transition, the critical theory exhibits an emergent (non-local) symmetry. The excitations at such critical point, sometimes fractionalized, correspond to "dynamically percolated" boundary excitations of the non-trivial SPT on one side of the transition. (The last statement was conjectured in Ref. [2].)

Most of the remaining of the main text, namely, Sections 2–4 present a sketch of the proof for the theorem. In these discussions we shall focus on physical arguments while keeping mathematics to a minimum level. The formal proofs are left in the appendices. The mathematical tool we use in this paper is the standard group cohomology cocycle manipulation. In the following we give the outline for the main text and appendices separately.

### *1.1. The outline of the main text*

In Section 2 we discuss the special $G \times Z_2^T$ SPT whose boundary, in the presence of $Z_2^T$ symmetry breaking field, exhibits the phase transition between a trivial and non-trivial $G$-symmetric SPTs *in one space dimension lower*. In Section 3 we discuss the NDSC condition imposed on the non-trivial $G$-symmetric SPTs on one side of the transition. In Section 4 we discuss the three possible scenarios (Fig. 2) of the SPT transition and relate them to the boundary physics of the $G \times Z_2^T$ SPT. In Section 5 we present simple examples of lattice models in one and two dimensions. These models are constructed under the framework enabled by the theorem. We shall discuss the phase transitions they exhibit. Finally, in Section 6 we conclude and discuss directions for future studies.

### 1.2. The outline of the appendices

In Appendix A we show how to construct the (fixed point) ground state wavefunction and their associated exactly solvable Hamiltonian for $\mathcal{G}$-symmetric SPTs in general dimensions. Here $\mathcal{G}$ can contain both unitary and anti-unitary elements. In Appendix B we construct the basis states spanning the low energy Hilbert space for the boundary of a $\mathcal{G}$-symmetric SPT, and derive how do they transform under the action of $\mathcal{G}$. In Appendix C we focus on $\mathcal{G} = G \times Z_2^T$ and dimension $= d + 1$. In (C.1) we focus on a particular subset of the cocycles of $H^{d+2}(G \times Z_2^T, U(1))$. In (C.2) we determine the condition for the non-trivialness of the chosen cocycles. In (C.3) we show that the SPTs constructed from these cocycles correspond to decorating the proliferated $Z_2^T$ domain walls with $G$-symmetric SPTs. In Appendix D we show that the boundary Hilbert space of the $G \times Z_2^T$ SPT contains an invariant subspace which is spanned by a basis isomorphic to the usual basis for studying $G$-symmetric SPTs in $d$ dimension. In part (D.1) we show how to utilize this basis to write down a family of $d$-dimensional lattice models exhibiting phase transition(s) between two inequivalent $G$-symmetric SPTs. For these models we show that the extra $Z_2^T$ symmetry acts non-locally. In Appendix E we show how the extra $Z_2^T$ symmetry implies there is no local $G \times Z_2^T$ symmetric Hamiltonian that can gap out the $d$-dimensional system without spontaneous symmetry breaking. In Appendices F and H we show how the framework developed in the paper can be applied to obtain simple lattice Hamiltonians in one and two space dimensions.

## 2. The $G \times Z_2^T$ symmetric SPT in $d + 1$ dimensions from proliferating decorated $Z_2^T$ domain walls

Generalizing the work of Ref. [6], we consider a *subset* of $(d + 1)$-dimensional $G \times Z_2^T$ symmetric SPTs constructed by proliferating $Z_2^T$ domain walls each "decorated" with a non-trivial $d$-dimensional $G$-symmetric SPT (satisfying the NDSC). The basis states spanning the Hilbert space for this problem is $\prod_i |\rho_i, g_i\rangle$ where $i$ labels the lattice sites and $\rho_i = \pm 1 \in Z_2^T$, $g_i \in G$. Hence each site has an Ising-like variable. This variable reverses sign under the action of $Z_2^T$. A state with non-zero expectation value of such Ising variable breaks the $Z_2^T$ symmetry. From such a symmetry breaking state we can construct a $Z_2^T$-symmetric state by "proliferating" the domain walls separating regions with opposite value of the Ising variable. (This means the ground state is a superposition of all possible Ising configurations.) Such domain walls are orientable $d$-dimensional manifolds and we choose the orientation consistently. To construct the $(d + 1)$-dimensional SPT, these domain walls are decorated with the $G$-symmetric $SPT_1$ or $\overline{SPT_1}$ (the inverse of $SPT_1$) according to the following rule. If the orientation of a domain wall points from the $+1$ domain to the $-1$ domain it is decorated with $SPT_1$. If the reverse is true it is decorated with $\overline{SPT_1}$. Because the $Z_2^T$ operation reverses the sign of the Ising variable, it must transforms $SPT_1$ into $\overline{SPT_1}$. A domain wall decorated with $\overline{SPT_1}$ is said to be conjugate to the one decorated with $SPT_1$ because when they are stacked together their respective SPTs combine to become trivial.

If we construct the wavefunctions for $SPT_1$ and $\overline{SPT_1}$ according to Appendix A, the wavefunction associated with $\overline{SPT_1}$ is the complex conjugate of that of $SPT_1$. Hence *the non-trivial element of $Z_2^T$ has two effects – it inverts the sign of the Ising variable as well as performing*

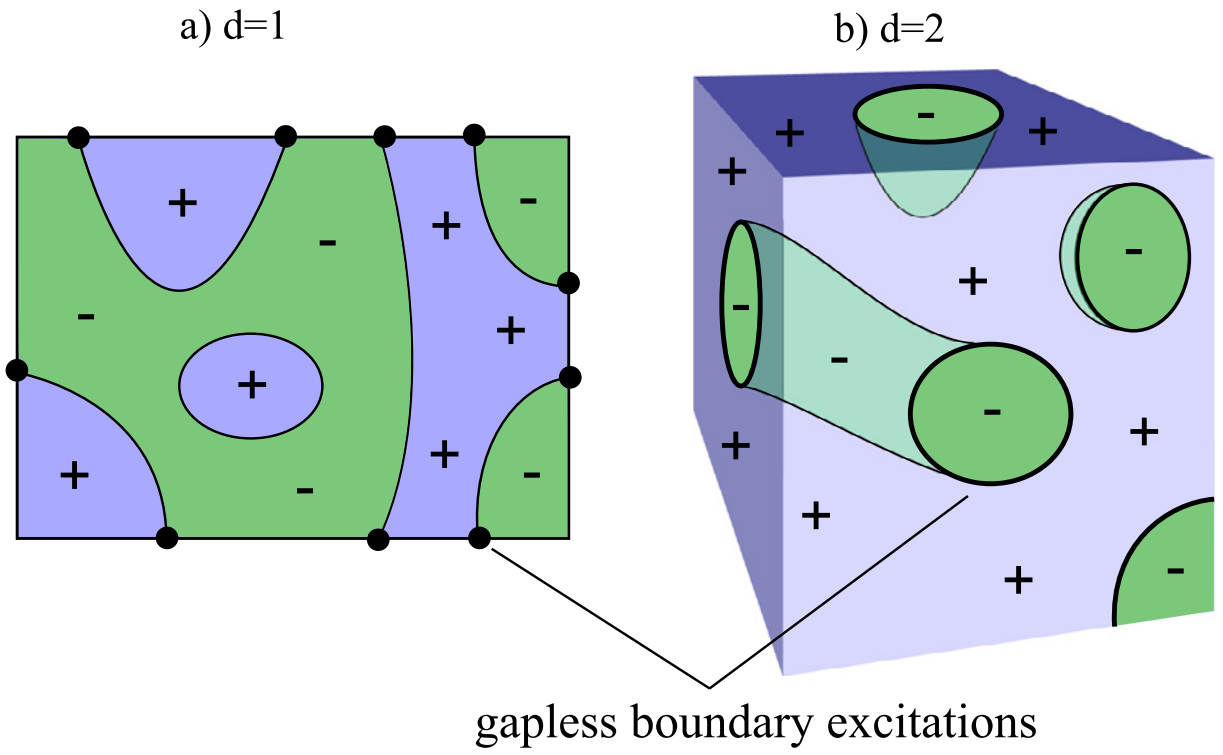

Fig. 3. (Color online.) Intersection of domain walls with the boundary of a $(d+1)$-dimensional system, for (a) $d=1$ and (b) $d=2$. The value of the $Z_2^T$ Ising variable for regions colored blue and green are $+1$ and $-1$ respectively. The domain walls are decorated with a $d$-dimensional SPT. Their intersections with the boundary are $(d-1)$-dimensional, denoted by black dots in (a) and solid lines in (b) respectively. These intersections host gapless boundary excitations of the SPT living on the domain walls.

*the complex conjugation on the wavefunction.* Because the Ising variable in question does not have to be time-reversal odd, the $Z_2^T$ discussed here can be different from the usual time reversal symmetry.

If the $(d+1)$-dimensional system has boundary, and which respects the $Z_2^T$ symmetry, the proliferated fluctuating bulk domain walls can intersect it. The intersection is $(d-1)$-dimensional (see Fig. 3) and is itself the boundary of the domain wall. Thus they harbor gapless boundary excitations of the SPT on the domain wall. However, when two "conjugate" intersections come close the gapless excitations on them can quantum tunnel. (A pair of conjugate intersections are the respective intersections of a pair of conjugate domain walls with the boundary.) When such quantum tunneling is strong a gap can open and effectively the two conjugate intersections annihilate each other.

## 3. The NDSC and the non-trivialness of the $G \times Z_2^T$-symmetric SPT

In Appendix (C.2) we prove mathematically that the state arises from proliferating the decorated $Z_2^T$ domain walls is non-trivial only if the SPT on the wall satisfies the NDSC. Now we explain why this condition is necessary. Let's suppose $SPT_1$, the SPT that the domain walls are decorated with, violates the NDSC and $SPT_1 = (SPT_{1/2})^2$ for certain $G$-symmetric $SPT_{1/2}$. In the following we show it is possible to perturb the boundary with a local $G \times Z_2^T$ symmetric Hamiltonian $\Delta H$ and gap out the gapless excitations.

Let $\Delta H$ coats the boundary with an additional layer of a $\overline{SPT_{1/2}}$ or $SPT_{1/2}$ depending on whether the $Z_2^T$ variable on the boundary is $+1$ or $-1$. Since $SPT_{1/2}$ is $G$-symmetric, $\Delta H$ respects the $G$-symmetry. Moreover because the coating switches from $SPT_{1/2}$ to $\overline{SPT_{1/2}}$ when the $Z_2^T$ variable is flipped, $\Delta H$ also respects the $Z_2^T$ symmetry. The fact $\Delta H$ is local is because the coating only depends on the value of the $Z_2^T$ variable locally.

Without loss of generality let's suppose the orientation of the domain wall points from the $+1$ domain to the $-1$ domain. This will induce an orientation on the intersection of the domain wall and the boundary. In Fig. 3 this means the blue region is "inside" and the green is "outside". Also let us choose the orientation of the coated film so that the orientation of the boundary be-tween $SPT_{1/2}$ and $\overline{SPT_{1/2}}$ agrees with that of the domain wall intersection. Without the coating the domain intersection carries the boundary gapless excitations of $SPT_1$. After the coating the interface between $SPT_{1/2}$ and $\overline{SPT_{1/2}}$ will be stacked on top of the original intersection. In the coated film of Fig. 4, when viewed from the $\overline{SPT_{1/2}}$ domain, the interface should host the boundary modes of $\overline{SPT_{1/2}}$. On the other hand when viewed from the $SPT_{1/2}$ domain the interface has the opposite orientation, thus it should host the conjugate of the $SPT_{1/2}$, i.e., the $\overline{SPT_{1/2}}$ boundary modes. As a result the stacked intersection/interface hosts the stacked boundary modes of $SPT_1$ and $\overline{SPT_{1/2}}^2 = \overline{SPT_1}$. Therefore they cancel and the gapless excitation on the domain wall/boundary intersection are gapped out. This means the $G \times Z_2^T$-symmetric SPT must be trivial because it is possible to add totally symmetric boundary perturbation to remove the gapless excitations. Hence in order for the SPT derived from proliferating the decorated $Z_2^T$ domain wall to be non-trivial the NDSC must be satisfied.

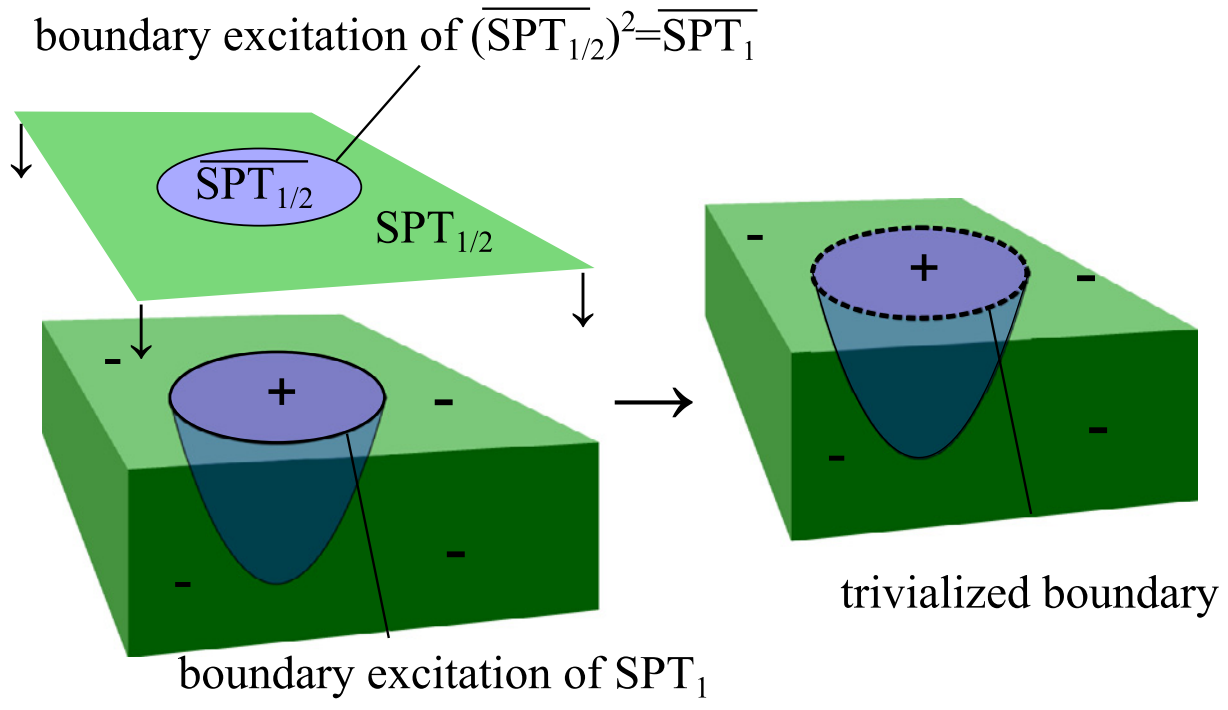

Fig. 4. (Color online.) Getting rid of the gapless boundary excitations if the SPT ($SPT_1$) used to decorate domain walls can be written as the square of another SPT ($SPT_{1/2}$). This is achieved by coating the surface with a layer of ($SPT_{1/2}$) on $-1$ domains and $\overline{SPT_{1/2}}$ on $+1$ domains (left panel). The combined boundary excitations on the intersection is gapped as denoted by the dashed line in the right panel.

## 4. The $Z_2^T$ symmetry breaking field and the three possible phase transition scenarios

Now let's assume the proliferated domain walls are decorated with the SPT satisfying the NDSC. In Appendix D we show that the boundary of such $G \times Z_2^T$ SPT has an invariant subspace "transplantable" to one dimension lower [7]. This invariant subspace can be made into the lowest-energy subspace by turning on *fully* $G \times Z_2^T$ symmetric *boundary* perturbation. The basis set of such subspace is $\prod_\mu |g_\mu\rangle_B$, $g_\mu \in G$ where $\mu$ labels the boundary sites. They transform under $G$ and the non-trivial element of $Z_2^T$ according to

$$S_g \prod_\mu |g_\mu\rangle_B = \prod_\mu |gg_\mu\rangle_B, \quad g \in G \tag{1}$$

$$S_{-1} \prod_\mu |g_\mu\rangle_B = \phi(\{g_\mu\}) K \prod_\mu |g_\mu\rangle_B, \quad -1 \in Z_2^T \tag{2}$$

where the pure phase

$$\phi(\{g_\mu\}) = \prod_\Delta [\nu_{d+1}(e, \{g_\mu\}_\Delta)]^{\sigma(\Delta)} \tag{3}$$

is the ground state wavefunction of the $G$-symmetric SPT used to decorate the domain wall, and $K$ stands for complex conjugation. Here $\nu_{d+1}$ is a $U(1)$ phase factor whose arguments are $d+2$ elements in $G$. It is a representative in the group cohomology class of $G$ that corresponds to the $G$-symmetric SPT, so it is fully determined by the group structure and the choice of a particular $G$-symmetric SPT (see Appendix A for a review of group cohomology). The product is carried over the $d$-dimensional simplices $\Delta$ which triangulate the $d$-dimensional boundary. $\sigma(\Delta) = \pm 1$ is the orientation of each simplex and $\{g_\mu\}_\Delta$ is a shorthand for the $d+1$ group elements assigned to the vertices of $\Delta$.

In Appendix D.1 we show how to construct a family of $d$-dimensional lattice models using the above basis set. These models depend on a parameter $\lambda \in [0, 1]$,

$$H(\lambda) = (1-\lambda)H_0 + \lambda H_1, \tag{4}$$

where

$$H_0 = -J \sum_\mu \sum_{g_\mu, g'_\mu} |\{g'_\mu\}\rangle_B \, {}_B\langle\{g_\mu\}|, \tag{5}$$

and

$$H_1 = -J \sum_\mu \sum_{g_\mu, g'_\mu} \frac{\phi(\{g'_\mu\})}{\phi(\{g_\mu\})} \overline{|\{g'_\mu\}\rangle_B} \, \overline{{}_B\langle\{g_\mu\}|}. \tag{6}$$

In the above equations $J > 0$ (and can be taken to very large value) and $\overline{|\{g'_\mu\}\rangle_B}$ stands for the complex conjugation of $|\{g'_\mu\}\rangle_B$. It is shown in Appendix C that both $H_0$ and $H_1$ are invariant under the action of $G$, and that the ground state of $H_0$ is the trivial $G$-symmetric SPT while the ground state of $H_1$ is the non-trivial $G$-symmetric SPT described by the wavefunction in Eq. (3). Upon the action of the non-trivial element of $Z_2^T$ transforms $H_0$ and $H_1$ according to

$$S_{-1} H_0 S_{-1}^{-1} = H_1 \text{ and } S_{-1} H_1 S_{-1}^{-1} = H_0. \tag{7}$$

Consequently $H(\lambda = 1/2)$ has an extra $Z_2^T$ symmetry (Eq. (2)). For other values of $\lambda$ there is only the $G$ symmetry (Eq. (1)). (Therefore we can view $\lambda - 1/2$ as a $Z_2^T$ symmetry breaking field.)

In Appendix E we prove that due to the non-local action of $S_{-1}$, $H(\lambda = 1/2)$ is either gapless or the $G \times Z_2^T$ symmetry is spontaneously broken. This implies at $\lambda = 1/2$ the $d$-dimensional system can be in one of the three following phases. (1) Gapless and $G \times Z_2^T$ symmetric. (2) Gapped but spontaneously breaks the $Z_2^T$ symmetry. (3) Gapped and spontaneously breaks the $G$ (or both the $G$ and $Z_2^T$) symmetry. Because at $\lambda = 1/2$ the system must be in one of the three phases discussed above, there are three possible routes for the phase transition from the trivial to non-trivial $G$-symmetry SPTs (Fig. 2). We discuss these three scenarios in the following. We shall do so from the view point of the $d$-dimensional system or that of the boundary of the $(d+1)$-dimensional system interchangeably.

### *4.1. Continuous phase transition*

This scenario corresponds to the boundary of the $G \times Z_2^T$ SPT being gapless. Under such condition the gapless excitations on the intersections of the fluctuating bulk domain walls and the boundary gives rise to a gapless boundary. These gapless-modes-infested domain wall intersections quantum fluctuate and delocalize throughout the boundary of the $(d+1)$-dimensional system. This is the "dynamic percolation" picture conjectured in Ref. [2].

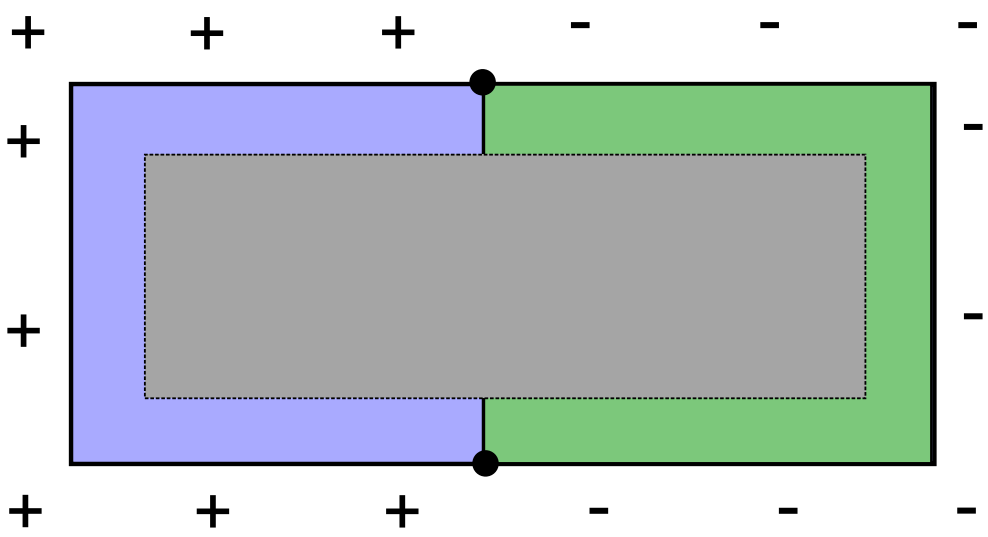

Fig. 5. (Color online.) The two inequivalent $G$-symmetric SPTs induced by opposite values of the boundary $Z_2^T$ symmetry breaking field. Here blue and green denote the trivial and non-trivial SPT, respectively. The interface between these two SPTs are also an $Z_2^T$ domain walls whose intersections with the boundary (the black dots) host the gapless boundary excitations of the SPT used to decorate the domain wall. The grey region in the center denotes the $G \times Z_2^T$ SPT with unbroken $Z_2^T$ symmetry.

Now let's imagine introducing the $Z_2^T$ symmetry breaking field on the boundary (and only on the boundary). Now the $Z_2^T$ domain wall can no longer intersect the boundary at sufficiently low energies. As a result the boundary is gapped. The two possible directions of the $Z_2^T$ symmetry breaking field leads to two $G$-symmetric SPTs corresponding to the $Z_2^T$ variable having opposite expectation values. In the following we show that these two SPTs are topologically inequivalent.

To do that we just need to demonstrate the interface between the two SPTs is necessarily gapless. This can be achieved by breaking the $Z_2^T$ symmetry so that half of the boundary has positive and the other half has negative $Z_2^T$ symmetry breaking field. The interface between these two halves are $Z_2^T$ domain walls and they have to connect to the fluctuating domain wall in the bulk (see Fig. 5). Hence they host gapless excitations. This implies the two $G$-symmetric SPTs on the boundary induced by opposite $Z_2^T$-breaking field are indeed inequivalent.

### 4.2. *First order phase transition*

Here we consider the case when the state at $\lambda = 1/2$ spontaneously breaks the $Z_2^T$ symmetry. In this case there will be degenerate ground states corresponding to the $Z_2^T$ variable having opposite expectation values. An infinitesimal $Z_2^T$ symmetry breaking field will lift the degeneracy and result in uniquely gapped $G$-symmetric phases on either side of $\lambda = 1/2$. From the boundary point of view because the $Z_2^T$ symmetry is spontaneously broken the fluctuating domain walls no longer intersect the boundary at low energies. This removes the gapless excitations associated with the interaction. The same argument associated with Fig. 5 implies the two gapped $G$ symmetric phases induced by opposite value of the symmetry breaking field are topologically inequivalent. Thus we have two distinct $G$-symmetric SPTs whose energy crosses at the transition point – i.e. a first order phase transition has occurred. This is depicted as the second scenario in Fig. 2.

### 4.3. *An intermediate symmetry breaking phase*

In the third scenario the boundary of the $G \times Z_2^T$ symmetric SPT spontaneously breaks the $G$ (or both the $G$ and $Z_2^T$) symmetry. Because of the $G$ symmetry breaking the gapless excitations at the domain wall intersections are gapped out. From the point of view of the $d$-dimensional system the $Z_2^T$ symmetry breaking field, i.e., the perturbation induced by $\lambda$ deviating from $1/2$, is $G$-symmetric. Because of the existence of energy gap, infinitesimal symmetry breaking field can only act within the degenerate ground state manifold (i.e. the subspace spanned by the degenerate ground states). Because the $G$ symmetry is spontaneously broken such ground state manifold must carry a multi-dimensional irreducible representation of $G$. Since the $Z_2^T$ symmetry breaking field is $G$ symmetric, it should be proportional to the identity operator within the ground state manifold. Consequently for values of $\lambda$ in the immediate neighborhood of $1/2$ the ground states remain degenerate and the $G$ symmetry remains spontaneously broken. When $\lambda$ deviates sufficiently from $1/2$ the $G$-symmetry has to be restored at some point because the limiting states at $\lambda = 0$ and $\lambda = 1$ are $G$-symmetric. Thus two Landau-like $G$ symmetry restoring critical points must intervene at intermediate $\lambda$. This gives rise to the possibility depicted as scenario (3) in Fig. 2.

In Section 5 we construct a simple solvable models for which scenario (1) and (3) are realized. Scenario (2) is suggested to occur in a numerical study on 2D $Z_2$ SPT phase transition[8]. We have not encountered an example where topological ordered [9] state appears on the boundary as discussed in Refs. [10–17], though it would be interesting for future studies.

## 5. Example: phase transition between $Z_2 \times Z_2$-symmetric SPTs in $d = 1$

### 5.1. A solvable case in one dimension

In one dimension there are two inequivalent $Z_2 \times Z_2$-symmetric SPTs ($H^2(Z_2 \times Z_2, U(1)) = Z_2$). We follow the recipe in Appendix C to construct the solvable Hamiltonians for the trivial and non-trivial $Z_2 \times Z_2$-symmetric SPTs and a family of interpolating Hamiltonians which realize scenario (1) of Fig. 2. Consider two spin-1/2 variables $\sigma_{2i-1}$ and $\sigma_{2i}$ in each unit cell $i$. The $Z_2 \times Z_2$ group acts as global $\pi$ rotations along $x$ and $z$ directions on all spins. As detailed in Appendix F the trivial/non-trivial Hamiltonians and the non-trivial element of $Z_2^T$ transformation are given by

$$H_0 = \sum_i (\sigma^x_{2i-1}\sigma^x_{2i} + \sigma^z_{2i-1}\sigma^z_{2i}) \tag{8}$$

$$H_1 = \sum_i (\sigma^x_{2i-2}\sigma^x_{2i-1} + \sigma^z_{2i}\sigma^z_{2i+1}) \tag{9}$$

$$H(\lambda) = (1-\lambda)H_0 + \lambda H_1 \tag{10}$$

$$S_{-1} = \prod_i \left(\frac{1+\sigma^z_{2i-1}\sigma^z_{2i+1}}{2} + \frac{1-\sigma^z_{2i-1}\sigma^z_{2i+1}}{2}\sigma^x_{2i}\right)K \tag{11}$$

Here $K$ stands for complex conjugation. It is straightforward to show that the trivial/non-trivial Hamiltonians transform into each other under $S_{-1}$. Eq. (8) and Eq. (9) and any linear combination of them are exactly solvable by going to the Majorana fermion representation (see Fig. F.1). In such representation $H_0$ contains intra-unit-cell (the rectangle boxes) coupling and $H_1$ contains inter-unit-cell coupling. The critical Hamiltonian $(H_0 + H_1)/2$ consists of two decoupled critical Majorana chains. As a result it exhibits central charge $c = 1$. In the spin-1/2 representation $(H_0 + H_1)/2$ is the XX model which possesses gapless spin-1/2 excitations. Since $H_1$, the dimerized XX model, has spin-1/2 edge states, this gives an explicit example where the gapless excitations of the critical state are delocalized, or dynamically percolated, edge excitations.

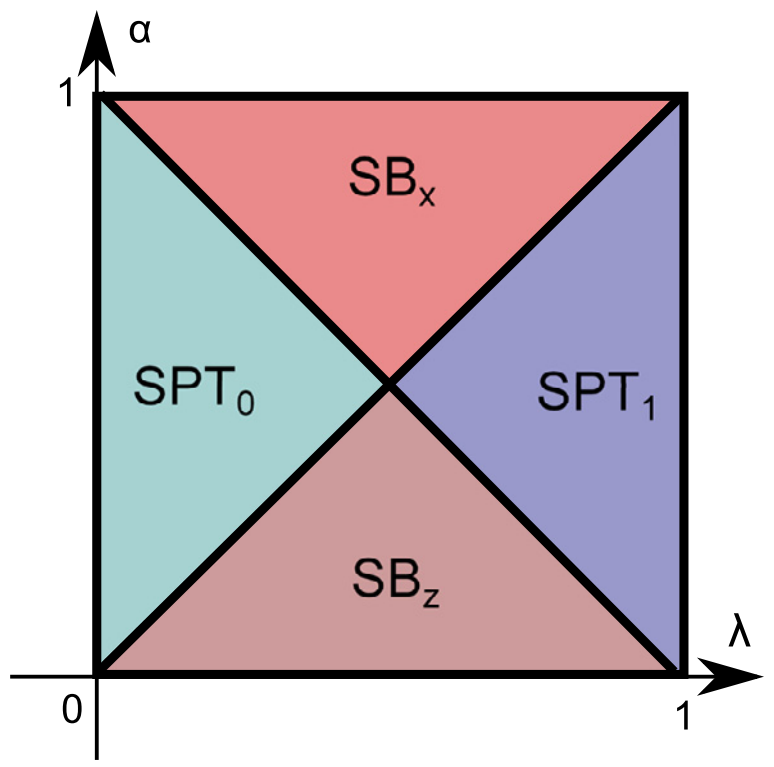

Fig. 6. (Color online.) The phase diagram of Eq. (12). The regions $SPT_0$, $SPT_1$ correspond to trivial and non-trivial SPTs, respectively. $SB_x$, $SB_z$ correspond to spontaneous symmetry-breaking with $\langle\sigma_x\rangle$ and $\langle\sigma_x\rangle$ non-zero, respectively. The solid black lines mark continuous phase transitions. Along the $\lambda = 1/2$ line there is either spontaneous symmetry breaking or gapless excitation.

Although the Hamiltonians in Eq. (10) is exactly solvable it has one undesirable feature, namely, it actually has higher symmetry ($U(1)$) than $Z_2 \times Z_2$ ($U(1) \times Z_2^T$ rather than $Z_2 \times Z_2 \times Z_2^T$ in the case of $\lambda = 1/2$). In the following we add perturbations to Eq. (10) to break the extra symmetry while maintain the solvability. Consider the following Hamiltonian

$$H(\lambda, \alpha) = \sum_i (1-\lambda)\left[\alpha\sigma^x_{2i-1}\sigma^x_{2i} + (1-\alpha)\sigma^z_{2i-1}\sigma^z_{2i}\right] + \lambda\left[\alpha\sigma^x_{2i-2}\sigma^x_{2i-1} + (1-\alpha)\sigma^z_{2i}\sigma^z_{2i+1}\right], \tag{12}$$

where $\alpha, \lambda \in [0, 1]$. For $\alpha \neq 1/2$ the symmetry of the model is reduced to $Z_2 \times Z_2$. Like Eq. (10) this model is exactly solvable after going to the Majorana basis. The phase diagram is shown in Fig. 6. Under $S_{-1}$, $\lambda$ transforms into $1-\lambda$ while $\alpha$ remains fixed. Along the line $(\lambda, \alpha) = (1/2, \alpha)$ the $Z_2^T$ symmetry is respected. Under that condition the system is either gapless or exhibits spontaneous symmetry breaking as predicted by the theorem in Appendix E. Interestingly, the critical point between SPT$_0$ and SPT$_1$ at $(\lambda, \alpha) = (1/2, 1/2)$ is also the transition point between two symmetry breaking phases. Moreover because the residual symmetry groups respected by the two symmetry breaking phases do not have subgroup relationship, the transition is an example of Landau forbidden transitions. Hence in this example the critical point between two SPTs is simultaneously the critical point of a Landau forbidden transition. Along the line $(\lambda, \alpha_0)$ where $\alpha_0 \neq 1/2$, the two SPT phases are intervened by a spontaneous symmetry breaking phase hence realizing the third scenario discussed in Section 4.

### 5.2. *Continuous phase transition in models with only $Z_2 \times Z_2^T$ symmetry (except at the critical point)*

In the last subsection after the removal of the extra $U(1)$ symmetry the transition between the SPTs is no longer direct. As a result one might wonder whether the continuous critical point is realizable without enlarging the symmetry group.

This motivates us to look for models with *only* $Z_2 \times Z_2$ symmetry (except at the critical point) while exhibiting a continuous direct transition between two inequivalent SPTs. The more general model is still exactly solvable at two limits, namely $\lambda = 0$ and 1 where it gives two inequivalent SPTs. However, unlike the simple example the model is not solvable for intermediate values of λ. In the following we perform density matrix renormalization group (DMRG) [18] calculation to study the intermediate λ including the critical point.

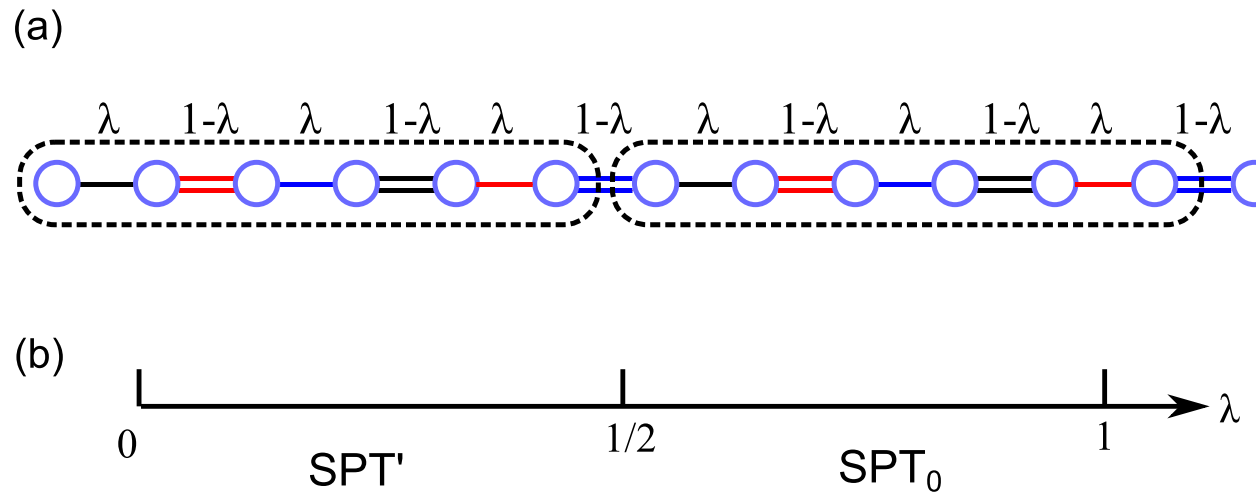

Fig. 7. (Color online.) (a) Sketch of the interactions in the model Hamiltonian of Eq. (13). Three different types of the interaction are represented by three different colored bonds. For example, black bonds denote $(S_i^x S_{i+1}^x + bS_i^y S_{i+1}^y)$, red bonds denote $(S_i^y S_{i+1}^y + bS_i^z S_{i+1}^z)$, and blue bonds denote $(S_i^z S_{i+1}^z + bS_i^x S_{i+1}^x)$. λ and $(1-\lambda)$ are the strength of the interactions. It is represented by single bonds and double bonds respectively. A dashed box denotes one unit cell. (b) Phase diagram for Eq. (13). The $\lambda < 1/2$ region is occupied by a non-trivial SPT, while the $\lambda > 1/2$ region is occupied by trivial SPT.

The Hamiltonian we consider is given by (as shown in Fig. 7(a))

$$\begin{aligned} H = \sum_{i=1}^{N/6} [\lambda(S_{6i-5}^x S_{6i-4}^x + bS_{6i-5}^y S_{6i-4}^y) \\ + (1-\lambda)(S_{6i-4}^y S_{6i-3}^y + bS_{6i-4}^z S_{6i-3}^z)] \\ + \lambda(S_{6i-3}^z S_{6i-2}^z + bS_{6i-3}^x S_{6i-2}^x) \\ + (1-\lambda)(S_{6i-2}^x S_{6i-1}^x + bS_{6i-2}^y S_{6i-1}^y) \\ + \lambda(S_{6i-1}^y S_{6i}^y + bS_{6i-1}^z S_{6i}^z) \\ + (1-\lambda)(S_{6i}^z S_{6i+1}^z + bS_{6i}^x S_{6i+1}^x)], \end{aligned} \tag{13}$$

where $S^x$, $S^y$ and $S^z$ are spin-1/2 operators, λ and $b$ are coupling parameters. The unit cell of Eq. (13) contain 6 sites each possessing a spin 1/2. These six spin-1/2s in each unit cell add to form integer total spins. The $Z_2 \times Z_2$ group is generated by π rotation around any two, e.g., $x$, $y$, spin axes for all spins.

When $b = 1$, the $Z_2^T$ transformation $S_{-1}$ flips $\lambda \leftrightarrow 1-\lambda$, is defined by $S_{-1} = U_1 U_2 K$, where

$$U_1 = \prod_{i=1}^{N/3} U_{XY,3i-2} U_{YZ,3i-1} U_{ZX,3i}$$

$$U_{AB,j} = \left(\frac{1+\sigma_j^A}{2} + \frac{1-\sigma_j^A}{2}\sigma_{j+1}^B\right)\left(\frac{1+i\sigma_{j+1}^B}{\sqrt{2}}\right)$$

$$U_2 = \prod_{i=1}^{N/3} \sigma_{3i-2}^y \sigma_{3i-1}^z \sigma_{3i}^x$$

It may be checked that $S_{-1}^2 = 1$ and it commutes with global $Z_2 \times Z_2$ rotations generated by $\prod_i \sigma_i^x$ and $\prod_i \sigma_i^z$.

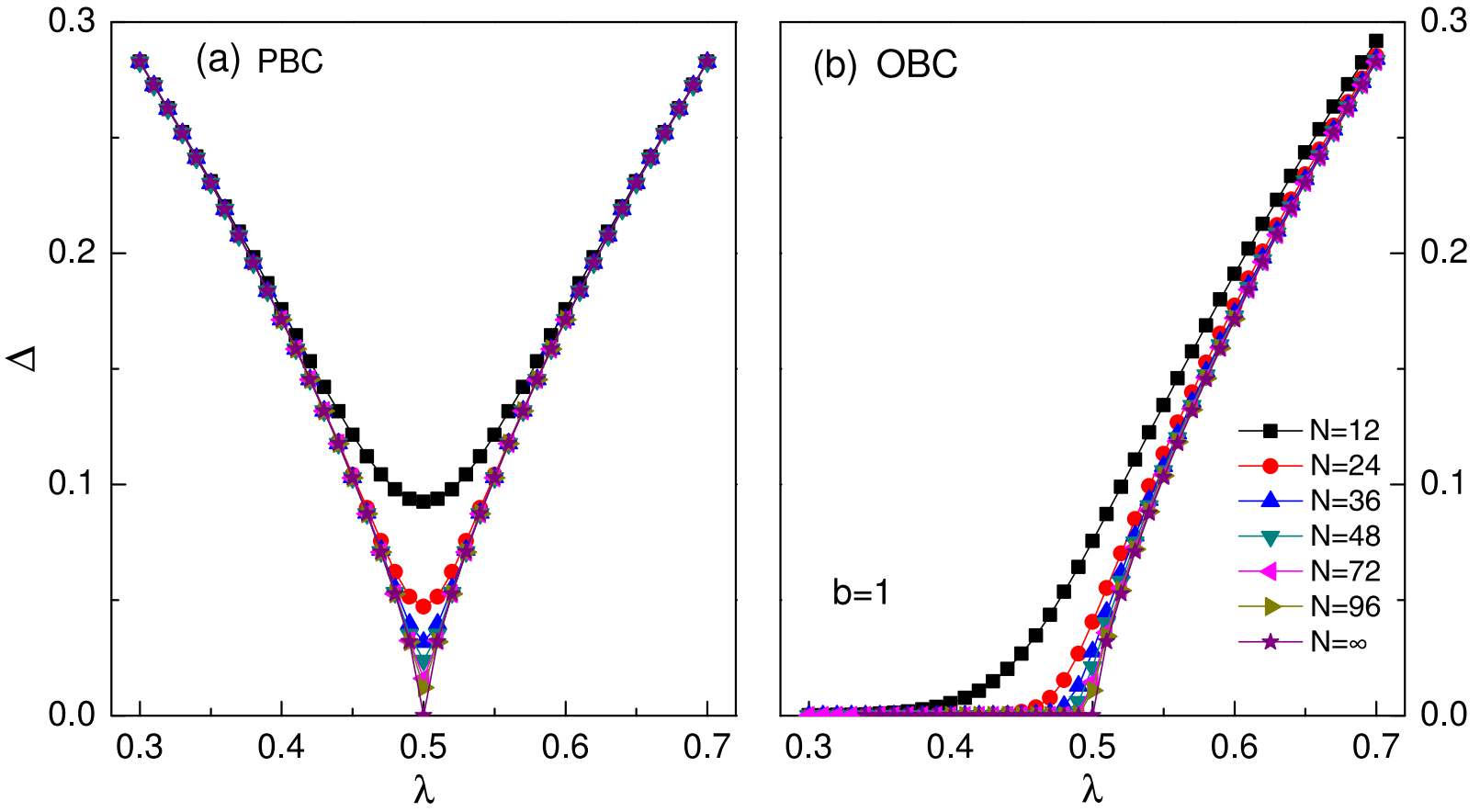

Fig. 8. (Color online.) Excitation gap $\Delta$ as a function of $\lambda$ at $b = 1$ (Eq. (13)). (a) For periodic boundary condition, $\Delta$ is finite except the critical point ($\lambda = 1/2$). (b) For open boundary condition, $\Delta = 0$ for $\lambda < 1/2$ due to the presence of gapless edge modes in the non-trivial SPT phase. For $\lambda > 1/2$ the SPT is trivial hence $\Delta > 0$.

In the limits $\lambda = 0$ and $\lambda = 1$ the system consists of decoupled dimers. It is simple to check that for $b > 0$ the ground state of each dimer is a spin singlet, and the bulk energy spectrum is gapped under periodic boundary condition. Under the open boundary condition there are gapless edge modes for $\lambda = 0$, while there is no edge state for $\lambda = 1$. So, these two limits are topologically distinct and we expect a phase transition between them for some intermediate value of $\lambda$.

The phase diagram of Eq. (13) is illustrated by Fig. 7(b). We find that for $\lambda < 0.5$ the system is in a non-trivial SPT phase. This is manifested by the fact that under periodic boundary condition (PBC) there is an energy gap while in open boundary condition (OBC) it is gapless (see Fig. 8). In contrast for $\lambda > 0.5$ the system is in a trivial SPT phase. This is manifested by the existence of an energy gap in both periodic and open boundary conditions (Fig. 8(a, b)). Interestingly, there is indeed a continuous phase transition between the two SPT phases occurring at $\lambda = 1/2$ for all $b > 0$ we have studied. Numerics indicate the central charge of this critical point is $c = 1$ (see Fig. 9), the same as that of the solvable case. Since $c = 1$ allows continuous varying critical exponents, we go on to extract the energy gap exponent $\alpha$,

$$\Delta \sim |\lambda - 1/2|^{\alpha}, \tag{14}$$

for different values of $b$. The results are shown in Fig. 10. For more details of the DMRG calculation see Appendix G.

The above example proves that scenario (1) in Fig. 2 is indeed attainable for phase transition between SPTs protected by *only* $Z_2 \times Z_2$.

### 5.3. Phase transition between $Z_2$ symmetric SPTs in 2D

In this subsection we follow the framework set in previous sections to construct a lattice model describing phase transition between by 2D $Z_2$-symmetric SPTs. (According to the cohomology group classification there are two inequivalent $Z_2$-symmetric SPTs in 2D.)

Consider a triangular lattice. For each site $i$ there is an Ising variable $\sigma_i := \sigma_i^z = \pm 1$. The trivial SPT Hamiltonian is

$$H_0 = -\sum_i \sigma_i^x. \tag{15}$$

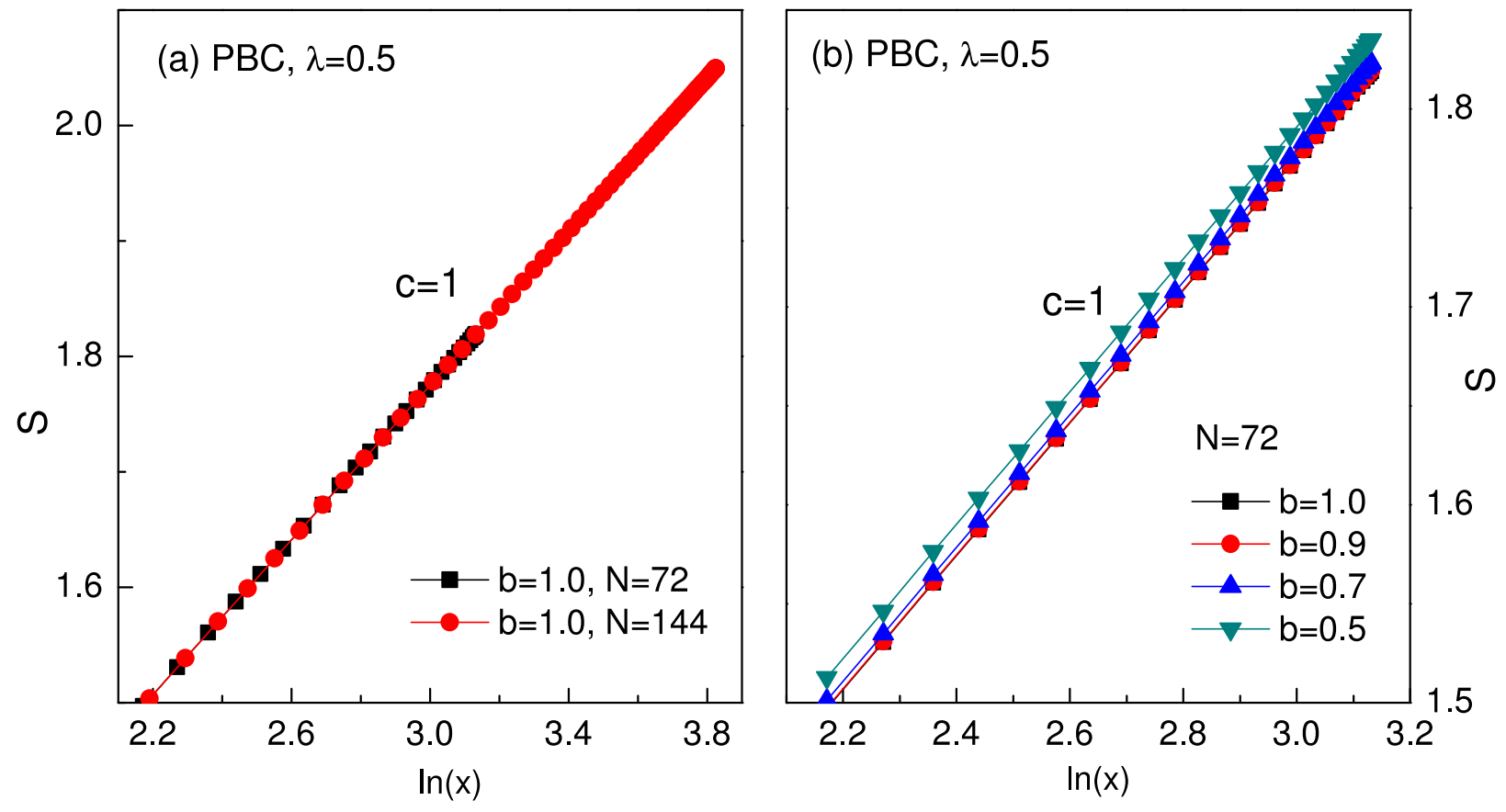

Fig. 9. (Color online.) Entanglement entropy scaling for Eq. (13) at $\lambda = 1/2$ and various values of $b$. Panel (a) shows the result for periodic $N = 72$ and 144 site chains for $b = 1$. Panel (b) shows the result for a periodic $N = 72$ site chain for various $b$ values. The fit to $S(x) = \frac{c}{3}\ln(x) + const$ extrapolates to a central charge $c = 1$. Here $x = \frac{N}{\pi}\sin(\frac{\pi l}{N})$, and $l$ is the subsystem length.

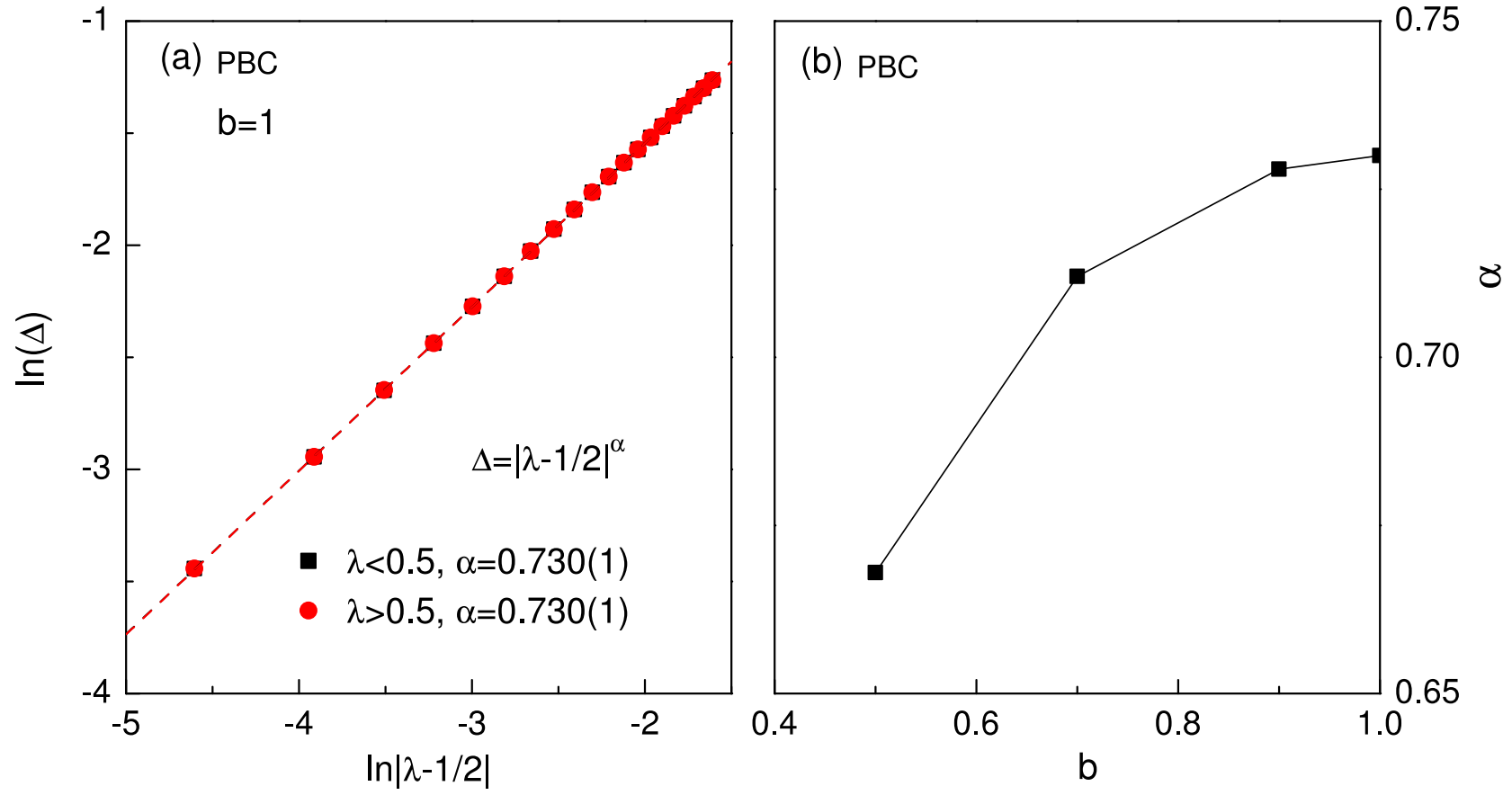

Fig. 10. (Color online.) (a) $\ln\Delta$ versus $\ln|\lambda - 1/2|$ for $b = 1$ under periodic boundary condition. Linearity implies $\Delta \sim |\lambda - 1/2|^\alpha$. (b) Gap exponent $\alpha$ for several values of $b$. Note that while $c = 1$ for all these $b$ values the gap exponent varies.

After some math the non-trivial SPT Hamiltonian can be reduced to

$$H_1 = \sum_i \left[ \Pi_{\langle j,k\rangle} i^{\left(\frac{1-\sigma_j\sigma_k}{2}\right)} \right] \left[ i^{(\sum_{j=1}^{6}\sigma_j)} \right] \sigma_i^x. \tag{16}$$

Here $\sigma_1, \ldots, \sigma_6$ designate the Ising variables on the six neighbors of $i$ as depicted in Fig. A.1(b), and the product $\Pi_{<j,k>}$ is performed over the six links connecting site $i$ and its six nearest neighbors. The non-trivial element of the $Z_2^T$ transformation is given by

$$S_{-1} = \prod_{\Delta}(-1)^{(\frac{1-\sigma_1}{2})(\frac{1+\sigma_2}{2})(\frac{1-\sigma_3}{2})} K \tag{17}$$

where $\sigma_1, \sigma_2, \sigma_3$ are the ordered vertices on each triangle $\Delta$. Again $S_{-1} H_0 S_{-1}^{-1} = H_1$ and $S_{-1} H_1 S_{-1}^{-1} = H_0$.

We construct the Hamiltonian to study the phase transition in exactly the same way as in Eq. (4). $H(\lambda)$ is solvable for $\lambda = 0$ and 1. For intermediate value of $\lambda$ it was suggested [8] numerically that there is a first-order transition at $\lambda = 1/2$. Thus scenario (2) in Fig. 2 is realized.

### 5.4. Phase transition between trivial and non-trivial phases of 1D integer spin chain

For integer spin chains $G = SO(3)$. The SPT phases are classified by $Z_2$, i.e., there is a trivial and a non-trivial SPT. For spin-1 chain, the non-trivial phase is also known as the Haldane [19] or the AKLT phase [20]. The continuum field theory describing the trivial and non-trivial phases is given by the following $O(3)$-non-linear sigma model (NLSM) with $\Theta = 0$ and $2\pi$, respectively

$$S = \frac{1}{2g}\int dxdt(\partial_\mu \hat{n})^2 + i\frac{\Theta}{4\pi}\int dxdt\ \hat{n}\cdot\partial_x\hat{n}\times\partial_t\hat{n}. \tag{18}$$

Here $\hat{n}$ is a 3-component unit vector. The critical point between the trivial and the non-trivial SPT is described by the $SU(2)_1$ Wess–Zumino–Witten (WZW) theory in $1+1$ dimensions [21]:

$$S = \frac{1}{2\tilde{g}}\int dxdt(\partial_\mu\hat{\Omega})^2 + \frac{i}{\pi}\int dxdt\int_0^1 du\ \epsilon^{abcd}\Omega_a\partial_x\Omega_b\partial_t\Omega_c\partial_u\Omega_d. \tag{19}$$

Here $\hat{\Omega}\in S^3$ is a 4-component unit vector, and $u$ is an extension parameter such that $\hat{\Omega}(u=0,x,t) = (0,0,0,1)$, and $\hat{\Omega}(u=1,x,t)$ is the physical $\hat{\Omega}(x,t)$. If the extra term $-\lambda\int dxdt\Omega_4(x,t)$ is added to Eq. (19), upon renormalization the low energy and long wavelength effective action flows to Eq. (18) with $\Theta = 0$ or $2\pi$ depending on the sign of $\lambda$. Hence $\lambda$ tunes the phase transition between the two SPTs. The emergent $Z_2^T$ symmetry discussed in this paper corresponds to reversing the sign of $\Omega_4$ followed by complex conjugation [22]. This symmetry is broken by the term $-\lambda\int dxdt\Omega_4(x,t)$. When $Z_2^T$ and $SO(3)$ (which rotates $\Omega_1$, $\Omega_2$, $\Omega_3$) symmetries are preserved, the $(1+1)$-dimensional boundary will either be gapless or degenerate [23].

## 6. Conclusion and discussion

In this paper we focus on the quantum phase transition between trivial and non-trivial symmetry protected topological states (SPTs) in $d$ dimensions. We prove that if the non-trivial SPT satisfies the "non-double-stacking condition" (see the theorem) all phase transition scenarios between them are captured by the boundary of a $(d+1)$-dimensional $G\times Z_2^T$ symmetric SPT in the presence of $Z_2^T$ symmetry breaking field. This result proves that at the critical point of the topological phase transition in question there is always emergent non-local symmetry. Moreover the symmetry operation associated with this non-local symmetry transforms one SPT phase into another. In addition our results provide explicit recipes for constructing $d$-dimensional lattice Hamiltonians describing different phase transition scenarios. As a byproduct we prove the conjecture made in Ref. [2], namely, the gapless excitations at the critical point between a trivial and non-trivial SPT consists of delocalized (or dynamically percolated) gapless boundary states of the non-trivial SPT. For future studies, we shall study how to describe phase transition between SPTs which do not satisfy the non-double-stacking condition. We will also consider the ramification of the interesting recent works which show the boundary of a three-dimensional SPT can exhibit topological order [10–17]. We ask what is the implication of this possibility on transitions between SPTs. Of course we are also interested in generating simple lattice models, especially in $d > 1$, describing the phase transition between SPTs, and in generalizing the approach here to the fermionic case.

## Acknowledgements

D.H.L. and Y.M.L. were supported by the U.S. Department of Energy, Office of Science, Basic Energy Sciences, Materials Sciences and Engineering Division, grant DE-AC02-05CH11231. H.C.J. was supported by the Department of Energy, Office of Science, Basic Energy Sciences, Materials Sciences and Engineering Division, under Contract DE-AC02-76SF00515. L.M.T. thanks Hong Yao and Ryan Thorngren for insightful discussions and is supported by the Department of Physics at UC Berkeley.

## Appendix A. The ground state wavefunction and exactly solvable bulk Hamiltonians from cocycles

We review the construction of bulk Hamiltonians and wavefunctions from cocycles [1,24]. In this section we focus on lattices residing on *closed* $d$-dimensional manifolds. A $n$-cochain with symmetry $\mathcal{G}$ is a map $c_n(g_0, g_1, \ldots, g_n) : \mathcal{G}^{n+1} \to U(1)$ which satisfies $c_n(gg_0, \ldots, gg_n) = c_n(g_0, \ldots, g_n)^\epsilon$, where $\epsilon = -1$ if $g$ is antiunitary and $+1$ for unitary $g$. A $n$-cocycle $\nu_n$ is a $n$-cochain which also satisfies the cocycle condition: $\partial \nu_n = 1$, where

$$(\partial \nu_n)(g_0, \ldots, g_{n+1}) = \prod_{i=0}^{n+1} \nu_n(g_0, \ldots, \hat{g}_i, \ldots, g_{n+1})^{(-1)^i}. \tag{A.1}$$

(Here $\hat{g}_i$ means $g_i$ is deleted.) If $\nu_n = \partial c_{n-1}$ for some $(n-1)$-cochain $c_{n-1}$ we say it is a coboundary. It may be checked that a coboundary also satisfies the cocycle condition, namely $\partial^2 c_{n-1} = 1$. Two cocycles related by the *multiplication* of a coboundary are viewed as equivalent.

$$\nu_n \sim \nu'_n = \nu_n \cdot \partial c_{n-1}. \tag{A.2}$$

The equivalence classes of $n$-cocycles form $H^n(\mathcal{G}, U(1))$ – the $n$th cohomology group. In Ref. [1] it is proposed that bosonic $\mathcal{G}$-symmetric SPTs in $d$ space dimensions are "classified" by $H^{d+1}(\mathcal{G}, U(1))$, i.e., each SPT is in one to one correspondence with a equivalence class of $(d+1)$-cocycles.

Suppose we have a triangulated $d$-dimensional *closed manifold* where vertices are the lattice sites. The Hilbert space for each site is spanned by $\{|g_i\rangle\}$ where $g_i \in \mathcal{G}$, and the total Hilbert space is spanned by the tensor product of the site basis, i.e., $|\{g_i\}\rangle = \prod_i |g_i\rangle$. The "fixed point" form (which is a particular representative) of the SPT states associated with the equivalence class of $\nu_{d+1}$ is equal to (Ref. [1] Section IX)

$$|\psi_0\rangle = \sum_{\{g_i\}} \phi(\{g_i\})\, |\{g_i\}\rangle$$

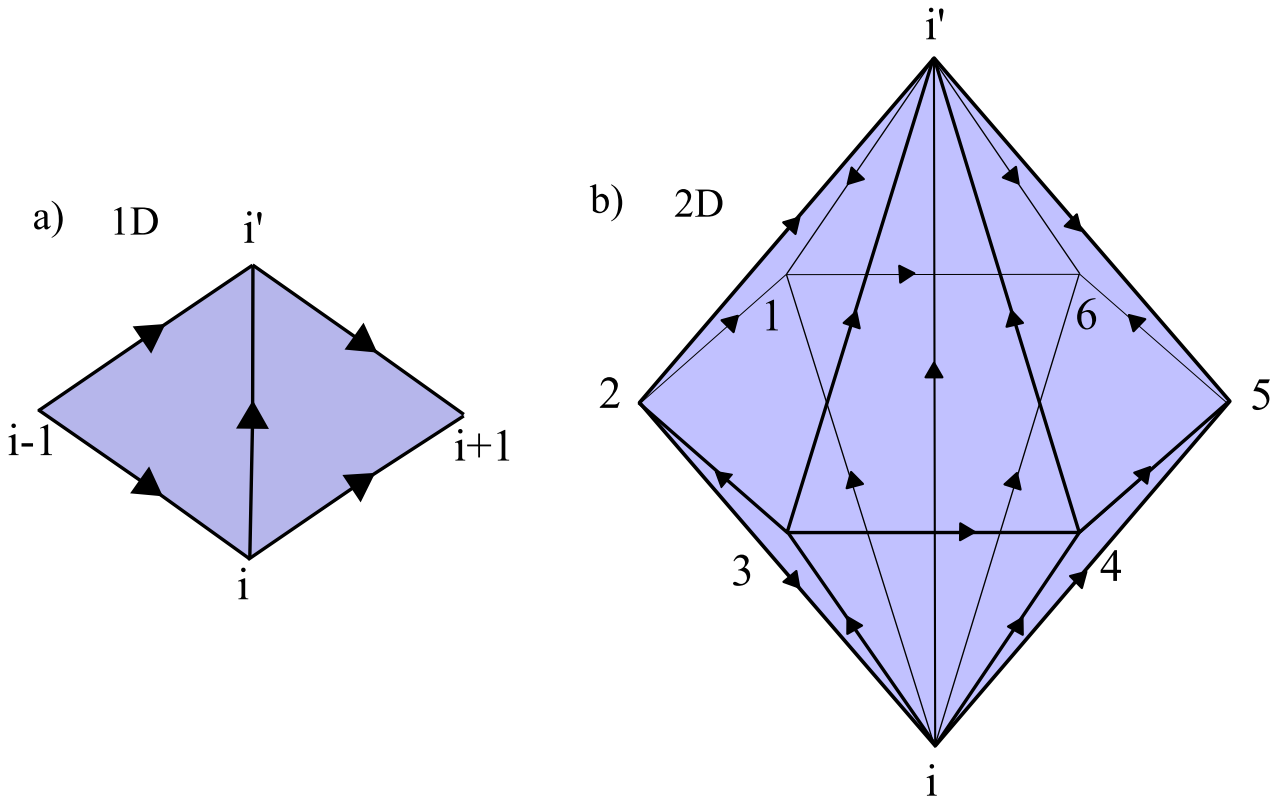

Fig. A.1. (Color online.) The construction of the exactly solvable bulk SPT Hamiltonians from cocycles. $B_i$ updates $g_i$ to $g'_i$. (a) For $d = 1$ the phase $\langle\{g'_i\}|B_i|\{g_i\}\rangle$ involves the cocycles associated with two triangles. (b) For $d = 2$ the phase $\langle\{g'_i\}|B_i|\{g_i\}\rangle$ involves the cocycles associated with six tetrahedrons.

$$\phi(\{g_i\}) = \prod_{\Delta}[\nu_{d+1}(e, \{g_i\}_\Delta)]^{\sigma(\Delta)}. \tag{A.3}$$

Here $e$ is the identity element, $\prod_\Delta$ is the product over the simplices of the triangulation, $\{g_i\}_\Delta$ is a shorthand for the $d + 1$ elements of $\mathcal{G}$ assigned to the ordered vertices of simplex $\Delta$, and $\sigma(\Delta) = \pm 1$ depending on the orientation of the simplex.

The Hamiltonian whose exact ground state is Eq. (A.3) is

$$H = -J\sum_i B_i, \tag{A.4}$$

where $J > 0$. The operator $B_i$ only affect the state on site $i$ and

$$\langle\{g'_k\}|B_i|\{g_k\}\rangle = \left(\prod_{k\neq i}\delta_{g'_k, g_k}\right)\frac{\phi(\{g'_k\})}{\phi(\{g_k\})}. \tag{A.5}$$

From Eqs. (A.3), (A.4) and (A.5) it is straightforward to show $B_i$ is a projection operator, $\mathrm{Tr}(B_i) = 1$ and $B_i|\psi_0\rangle = |\psi_0\rangle$. In addition using the cocycle condition it can be shown that $[B_i, B_j] = 0\ \forall i, j$. So $|\psi_0\rangle$ is the unique gapped ground state of $H$. In addition using the cocycle condition $\phi(g_1, \ldots, g'_i, \ldots g_N)/\phi(g_1, \ldots, g_i, \ldots g_N)$ can be shown to depend on the $g$'s in the neighborhood of site $i$, hence the Hamiltonian is local. Examples for 1D and 2D are given below.

In 1D $\phi(\{g'_i\})/\phi(\{g_i\})$ can be reduced via the cocycle condition into (Fig. A.1(a)):

$$\frac{\nu_2(g_{i-1}, g_i, g'_i)}{\nu_2(g_i, g'_i, g_{i+1})}$$

In 2D, suppose we have a triangular lattice, each site has six neighbors $1, \ldots, 6$. In this case $\phi(\{g'_i\})/\phi(\{g_i\})$ involves the $g$'s on six tetrahedrons (Fig. A.1(b)):

$$\frac{\nu_3(g_3, g_4, g_i, g'_i)\nu_3(g_4, g_i, g'_i, g_5)\nu_3(g_i, g'_i, g_5, g_6)}{\nu_3(g_3, g_2, g_i, g'_i)\nu_3(g_2, g_i, g'_i, g_1)\nu_3(g_i, g'_i, g_1, g_6)}$$

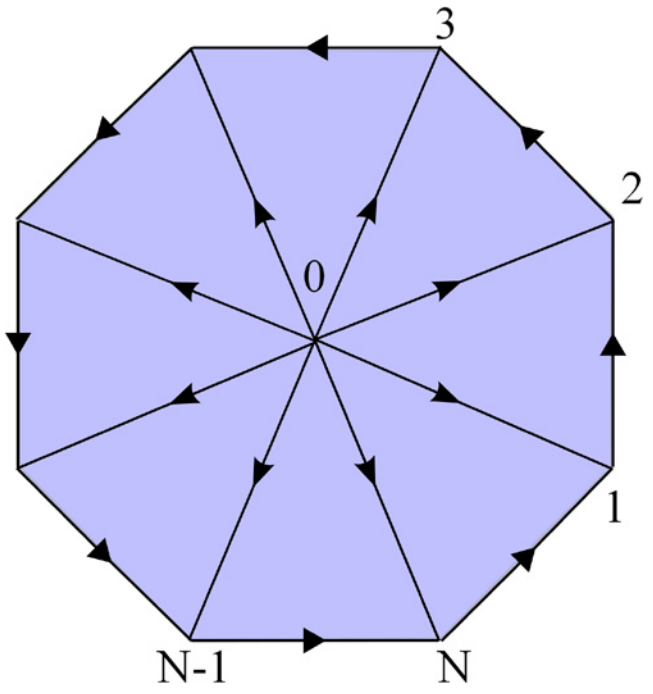

Fig. B.1. (Color online.) A single site 0 represents all the bulk degrees of freedom. We use a convention where arrows point from the bulk to the boundary.

## Appendix B. Boundary basis and their symmetry transformations

In this section we consider lattices on *open* $(d+1)$-dimensional manifolds. In general, a ground state wavefunction on an open manifold is defined subject to fixed boundary site configurations. Then the wavefunction is given by (A.3) summed over all bulk site configurations with the product carried out over simplices within the open manifold. In this way, if we have two open manifolds with the same boundary, we may take the direct product of their wavefunctions, identify their boundary sites and sum over all the possible boundary site configurations to recover the wavefunction on a closed manifold.

Let the vertices (sites) of the triangulated $d$-dimensional boundary be labeled by Greek index $\mu$, and let there be a single "bulk site" labeled by "0". Together with the boundary sites they triangulate a $(d+1)$-dimensional open manifold. Our convention for vertex ordering is that all arrows point from 0 to $\mu$ (see Fig. B.1). Note that the assumption of having a single bulk site is purely for ease of manipulation, and does not put constraint on the topology of the manifold considered. It can be checked that our final result, the boundary transformation (B.6) remains unchanged even when more sites are added in the bulk.

The total Hilbert space is spanned by $\{|g_0, \{g_\mu\}\rangle\}$, and using Eq. (A.3) (except the spatial dimension is $d+1$ rather than $d$) we write down the expression for the ground state subject to boundary conditions $\{g_\mu\}$ as discussed before

$$|\{g_\mu\}\rangle_B = \sum_{g_0}\prod_{\Delta}[\nu_{d+2}(e, g_0, \{g_\mu\}_\Delta)]^{\sigma(\Delta)}|g_0, \{g_\mu\}\rangle \tag{B.1}$$

Upon the action of the global symmetry group both the bulk and boundary states are transformed. Let $S_g$ be the representation of the symmetry operation $g \in G$, we have

$$S_g|g_0, \{g_\mu\}\rangle = |gg_0, \{gg_\mu\}\rangle, \tag{B.2}$$

and

$$\begin{aligned} S_g|\{g_\mu\}\rangle_B &= \sum_{g_0}\prod_{\Delta}[\nu_{d+2}(e, g_0, \{g_\mu\}_\Delta)]^{\epsilon\sigma(\Delta)}|gg_0, \{gg_\mu\}\rangle \\ &= \sum_{g_0}\prod_{\Delta}[\nu_{d+2}(g, gg_0, \{gg_\mu\}_\Delta)]^{\sigma(\Delta)}|gg_0, \{gg_\mu\}\rangle \end{aligned}$$

$$= \sum_{g_0} \prod_{\Delta} [\nu_{d+2}(g, g_0, \{gg_\mu\}_\Delta)]^{\sigma(\Delta)} |g_0, \{gg_\mu\}\rangle \tag{B.3}$$

Using the cocycle condition

$$\partial \nu_{d+2}(g, e, g_0, \{gg_\mu\}_\Delta) = 1 \tag{B.4}$$

the last line of Eq. (B.3) can be equated with

$$\begin{aligned} &\sum_{g_0} \prod_{\Delta} \Big\{ \nu_{d+2}(g, e, \{gg_\mu\}_\Delta)^{\sigma(\Delta)} \\ &\quad \times \left[ \prod_{i=0}^{d} \nu_{d+2}(g, e, g_0, \{gg_\mu\}_{\Delta_i})^{(-1)^{i+1}\sigma(\Delta)} \right] \times \nu_{d+2}(e, g_0, \{gg_\mu\}_\Delta)^{\sigma(\Delta)} \Big\} |g_0, \{gg_\mu\}\rangle \\ &= \left[ \prod_{\Delta} \nu_{d+2}(g, e, \{gg_\mu\}_\Delta)^{\sigma(\Delta)} \right] |\{gg_\mu\}\rangle_B \end{aligned} \tag{B.5}$$

In the second line $\Delta_i$ is the shorthand for the $(d-1)$-dimensional simplex which is a face of $\Delta$ obtained by removing its $i$th vertex. In the last step we have used the fact that each $(d-1)$-dimensional simplex is the face to two opposite orientation $d$-dimensional simplices $\Delta$'s, hence their contributions cancel in the product. Therefore

$$S_g|\{g_\mu\}\rangle_B = \left[ \prod_{\Delta} \nu_{d+2}(g, e, \{gg_\mu\}_\Delta)^{\sigma(\Delta)} \right] |\{gg_\mu\}\rangle_B \tag{B.6}$$

For example when $d = 0$, the edge of the 1D SPT are two points $g_L, g_R$. Under $g$ they transform as

$$S_g|g_L, g_R\rangle_B = \frac{\nu_2(g, e, gg_R)}{\nu_2(g, e, gg_L)} |gg_L, gg_R\rangle_B \tag{B.7}$$

The phase $\frac{\nu_2(g,e,gg_R)}{\nu_2(g,e,gg_L)}$ is interpreted as the edge states being carrying the projective representation of the symmetry group. For $d = 1$, the edge forms a ring labeled by $\mu = 1, \ldots, N$. In this case

$$S_g|\{g_\mu\}\rangle_B = \prod_{\mu=1}^{N} \nu_3(g, e, gg_\mu, gg_{\mu+1}) |\{gg_\mu\}\rangle_B \tag{B.8}$$

In Ref. [25], this transformation with $G = Z_2$ is an example of "Matrix Product Unitary Operator". They also show that if $\nu_3$ is non-trivial, then the edge states made up of linear combination of $|\{g_\mu\}\rangle_B$ cannot be a short ranged entangled state.

## Appendix C. Construction and the physical interpretation of $G \times Z_2^T$ SPTs

### C.1. The special subset of cocycles

In this section we use a particular cocycle of $H^{d+2}(G \times Z_2^T, U(1))$ to construct the SPT in $d+1$ dimensions. This cocycle is given by

$$\nu_{d+2}(\rho_0 g_0, \rho_1 g_1, \ldots, \rho_{d+2} g_{d+2}) = [\nu_{d+1}(g_1, \ldots, g_{d+2})]^{\frac{\rho_1 - \rho_0}{2}}, \tag{C.1}$$

where $g_i \in G$, $\rho_i = \pm 1 \in Z_2^T$ and $\nu_{d+1}$ is a non-trivial cocycle of $H^{d+1}(G, U(1))$. It is straightforward to verify that $\nu_{d+2}$ indeed satisfies the cocycle condition.

*C.2. The non-double-stacking condition on the $\nu_{d+1}$ in Eq. (C.1)*

In the following we prove that so long as $\nu_{d+1}$ cannot be written as the square of another $d+1$ cocycle, say, $\tilde{\nu}_{d+1}$, the $\nu_{d+2}$ given by Eq. (C.1) is a non-trivial cocycle of $H^{d+2}(G\times Z_2^T, U(1))$. More precisely we will show $\nu_{d+2}$ is trivial if and only if $\nu_{d+1}=(\tilde{\nu}_{d+1})^2\cdot\partial c'_d$ for some (not necessarily trivial) $G$-symmetric cocycle $\tilde{\nu}_{d+1}$ and cochain $c'_d$. Thus the higher-dimensional SPT constructed using Eq. (C.1) is trivial if and only if $\nu_{d+1}$ is a double-stacking of another $\tilde{\nu}_{d+1}$.

To prove the "if" part, suppose $\nu_{d+1}=(\tilde{\nu}_{d+1})^2\cdot\partial c'_d$. Substitute this into Eq. (C.1) we obtain

$$\nu_{d+2}(\rho_0 g_0,\rho_1 g_1,\ldots,\rho_{d+2},g_{d+2})=\tilde{\nu}_{d+1}(g_1,\ldots,g_{d+2})^{\rho_1-\rho_0}\partial c'_d(g_1,\ldots,g_{d+2})^{\frac{\rho_1-\rho_0}{2}}$$

If we define

$$c_{d+1}(\rho_1 g_1,\ldots,\rho_{d+2}g_{d+2}):=\tilde{\nu}_{d+1}(g_1,\ldots,g_{d+2})^{\rho_1}c'_d(g_2,\ldots,g_{d+2})^{\frac{\rho_1-\rho_2}{2}}$$

Then using the cocycle condition on $\tilde{\nu}_{d+1}$, it may be checked that $c_{d+1}$ is a $G\times Z_2^T$ cochain (hence is symmetric under the action of the group) and $\partial c_{d+1}=\nu_{d+2}$. So $\nu_{d+2}$ is a trivial cocycle.

To prove the "only if" part, suppose $\nu_{d+2}=\partial c_{d+1}$ for some $G\times Z_2^T$ cochain $c_{d+1}$, i.e.

$$\begin{aligned}&\nu_{d+2}(\rho_0 g_0,\ldots,\rho_{d+2}g_{d+2})\\&\quad=\partial c_{d+1}(\rho_0 g_0,\ldots,\rho_{d+2}g_{d+2})=c_{d+1}(\rho_1 g_1,\ldots,\rho_{d+2}g_{d+2})\\&\qquad\times\prod_{k=1}^{d+2}c_{d+1}^{(-1)^k}(\rho_0 g_0,\rho_1 g_1,\ldots,\widehat{\rho_k g_k},\ldots,\rho_{d+2}g_{d+2})\end{aligned}\tag{C.2}$$

We will prove the $c_{d+1}$ in question must satisfy

$$\partial c_{d+1}(g_0,\ldots,g_{d+2})=\partial c_{d+1}(\rho_0 g_0,\ldots,\rho_{d+2}g_{d+2})\Big|_{\rho_i\to 1\ \forall i}=1,\tag{C.3}$$

i.e., upon setting all $\rho_i=1$ the $c_{d+1}$ in question is a $G$-cocycle. In addition we shall prove that if $\nu_{d+2}=\partial c_{d+1}$ the $\nu_{d+1}$ in Eq. (C.1) must satisfy

$$\nu_{d+1}=c^2_{d+1}\Big|_{\rho_i\to 1\ \forall i}\cdot\partial c'_d\tag{C.4}$$

for certain $G$-coboundary $\partial c'_d$. (Assuming (C.3) and (C.4) hold, then we choose $\tilde{\nu}_{d+1}=c_{d+1}\Big|_{\rho_i\to 1\ \forall i}$ to complete the proof.)

To show (C.3), we first note that by taking $\rho_0=\rho_1=1$ in Eq. (C.1) and the second line of Eq. (C.2), we have

$$1=\partial c_{d+1}(g_0,g_1,\rho_2 g_2,\ldots,\rho_{d+2}g_{d+2})\tag{C.5}$$

Then (C.3) follows directly by further setting $\rho_i=1$ for all $i$.

To show (C.4), we let $\rho_0=-1$, $\rho_i=1$ for $i\neq 0$ and $g_0=g_1$ in (C.1) and (C.2). Then

$$\begin{aligned}\nu_{d+1}(g_1,\ldots,g_{d+2})&=c_{d+1}(g_1,\ldots,g_{d+2})\times\prod_{k=1}^{d+2}c_{d+1}^{(-1)^k}(-g_1,g_1,\ldots,\widehat{g_k},\ldots,g_{d+2})\\&=c_{d+1}\cdot\gamma_1\end{aligned}\tag{C.6}$$

where $\gamma_l$ is defined as follows. For $l=1,\ldots,d+2$,

$$\gamma_l(g_1,\dots,g_{d+2}) := \prod_{k=l}^{d+2} c_{d+1}^{(-1)^{l-k+1}}(-g_1,\dots,-g_l,g_l,\dots,\widehat{g_k},\dots,g_{d+2})$$

It turns out that $\gamma_l \sim \gamma_{l+1}$, for $l = 1,\dots,d+1$. The proof is as follows:

$$\begin{aligned}
\gamma_l &= c_{d+1}^{-1}(-g_1,\dots,-g_l,g_{l+1},\dots,g_{d+2}) \\
&\quad \times \prod_{k=l+1}^{d+2} c_{d+1}^{(-1)^{l-k+1}}(-g_1,\dots,-g_l,g_l,\dots,\widehat{g_k},\dots,g_{d+2}) \\
&= c_{d+1}^{-1}(-g_1,\dots,-g_l,g_{l+1},\dots,g_{d+2}) \\
&\quad \times \prod_{k=1}^{l} c_{d+1}^{(-1)^{l-k}}(-g_1,\dots,\widehat{-g_k},\dots,-g_{l+1},g_{l+1},\dots,g_{d+2}) \\
&\quad \times \partial c_d'^{(-1)^{l+1}}(g_1,\dots,g_{d+2}) \\
&\sim \prod_{k=1}^{l+1} c_{d+1}^{(-1)^{l-k}}(-g_1,\dots,\widehat{-g_k},\dots,-g_{l+1},g_{l+1},\dots,g_{d+2}) \\
&= \prod_{k=l+1}^{d+2} c_{d+1}^{(-1)^{l-k}}(-g_1,\dots,-g_{l+1},g_{l+1},\dots,\widehat{g_k},\dots,g_{d+2}) \\
&\quad \times \partial c_{d+1}^{(-1)^{l+1}}(-g_1,\dots,-g_{l+1},g_{l+1},\dots,g_{d+2}) \\
&= \gamma_{l+1}
\end{aligned}$$

where $c_d'$ is a $G$-symmetric cochain defined by:

$$c_d'(\tilde{g}_1,\dots,\tilde{g}_{d+1}) := c_{d+1}(-\tilde{g}_1,\dots,-\tilde{g}_l,\tilde{g}_l,\dots,\tilde{g}_{d+1})$$

and in the last line we used

$$\partial c_{d+1}(-g_1,\dots,-g_{l+1},g_{l+1},\dots,g_{d+2}) = 1$$

which follows from (C.5). Thus

$$\gamma_1 \sim \gamma_{d+2} = c_{d+1}^{-1}(-g_1,\dots,-g_{d+2}) = c_{d+1}(g_1,\dots,g_{d+2}). \tag{C.7}$$

In the last line we have used the fact that $c_{d+1}$ is a $G \times Z_2^T$ symmetric cochain. Substituting equation (C.7) into (C.6), (C.4) is proven.

### *C.3. Interpreting the wavefunction as decorated domain walls*

The wavefunction,

$$\psi(\{\rho_i g_i\}) = \prod_{\Delta}[\nu_{d+2}(e,\{\rho_i g_i\}_\Delta)]^{\sigma(\Delta)}, \tag{C.8}$$

constructed from (C.1) can be viewed as having time-reversal domain walls decorated with lower-dimensional SPT.

To demonstrate this, we first derive an alternative form for the ground state wavefunction discussed in Appendix A in general. In this subsection it is assumed the system is a closed manifold.

Suppose we have a $\mathcal{G}$ protected SPT in $d$ dimensions with associated cocycle $H^{d+1}(\mathcal{G}, U(1))$. Applying the cocycle condition $\partial\nu_{d+1}(e, \{g_i\}_\Delta, e) = 1$ on the ground state wavefunction (A.3), we obtain

$$\begin{aligned}\phi(\{g_i\}) &= \prod_\Delta [\nu_{d+1}(e, \{g_i\}_\Delta)]^{\sigma(\Delta)} \\ &= \prod_\Delta [\nu_{d+1}^{(-1)^{d+1}}(\{g_i\}_\Delta, e)]^{\sigma(\Delta)} \times \left(\prod_{j=0}^{d} \nu_{d+1}^{(-1)^{d+j}}(e, \{g_i\}_{\Delta_j}, e)\right)^{\sigma(\Delta)} \end{aligned} \tag{C.9}$$

The last factor is identity because each simplex $\Delta_i$ is the face to two simplices $\Delta$ whose contributions cancel. Therefore we may alternatively write the ground state wavefunction as

$$\phi(\{g_i\}) = \prod_\Delta \underbrace{[\nu_{d+1}(\{g_i\}_\Delta, e)]^{\sigma(\Delta)(-1)^{d+1}}}_{:=\phi_\Delta(\{g_i\})^{\sigma(\Delta)}} \tag{C.10}$$

Now with $\mathcal{G} = G \times Z_2^T$ and using Eq. (C.8), we have

$$\begin{aligned}\psi(\{\rho_i g_i\}) &= \prod_\Delta [\nu_{d+1}(\{g_i\}_{\Delta_0}, e)]^{\frac{\rho_{1(\Delta)}-\rho_{0(\Delta)}}{2}\sigma(\Delta)(-1)^{d+2}} \\ &= \prod_\Delta \left[\phi_{\Delta_0}(\{g_i\})\right]^{\sigma(\Delta)(\frac{\rho_{1(\Delta)}-\rho_{0(\Delta)}}{2})} \end{aligned} \tag{C.11}$$

Here $\Delta_0$ is the simplex $\Delta$ with the first of its ordered vertices deleted, and $0(\Delta)$, $1(\Delta)$ stand for the integers labeling the first and second vertices of the simplex $\Delta$.

We now assume the system is on a $(n = d + 1)$-dimensional torus $T^n$ with a specific triangulation defined as follows. Let the vertices form a simple hypercubic structure on $T^n$. Each hypercube is identified with the region $\{(x_1, \dots, x_n) : 1 \geq x_i \geq 0\}$. Now cut the hypercube into $n!$ simplices $\Delta(P)$, each la belled by a permutation $P$ of $(1, \dots, n)$. The simplex $\Delta(P)$ occupies the region with $1 \geq x_{P(1)} \geq \cdots \geq x_{P(n)} \geq 0$. There are $2^n$ vertices in the hypercube, each has all its coordinates equal to 0 or 1. For each vertex $v$, let $\mathcal{N}(v)$ be the number of 1's in its coordinate. The arrows defining the ordering of the vertices run from a vertex with a smaller $\mathcal{N}$ to a vertex with larger $\mathcal{N}$. In each simplex, there are exactly one vertex with any given $\mathcal{N}$, which ranges from 0 to $n$. One may check that such ordering of vertices is consistent on faces shared by two hypercubes. For every simplex, the lowest vertex is the origin $v_0$ with all coordinates zero. The second lowest vertex has exact one 1 in its coordinates, which we label $v_k$ such that its $j$-th component is $\delta_{kj}$. We also let $F_k$ to be the face on the hypercube whose vertices gas all their $k$-th coordinate = 1. $F_k$ is itself a $(n-1)$-dimensional hypercube. A domain wall between $v_0$ and $v_k$ lives in the hyperplane equidistant from $v_0$ and $v_k$, which we associate to $F_k$ via a translation of $(\frac{1}{2}, \dots, \frac{1}{2})$.

The orientation of simplex $\sigma(\Delta(P))$ is given by $\mathrm{sgn}(P)$. This induces an orientation on the face $\Delta(P)_0$ given by $\sigma(\Delta(P)_0) = \sigma(\Delta(P))$. Now the interpretation of (C.11) is that, if there is a domain wall between $v_0$ and $v_k$, then on $F_k$ there will live a lower-dimensional SPT with ground state wavefunction given by Eq. (C.10) or its complex conjugate depending on whether the orientation of $F_k$ points from the $\rho = +1$ vertex to the $\rho = -1$ vertex or vice versa.

An example is given in Fig. C.1. for the case where $d + 1 = 2$. Here the dashed red line is the actual domain wall and the solid red line is where the $d$-dimensional $G$-symmetric SPT resides. The wavefunction with $Z_2^T$ variables fixed as in Fig. C.1 is

$$
\begin{aligned}
\psi(\{g_i\}) &= \prod_{j=0}^{3} [\nu_2(g_{3j}, g_{3(j+1)}, e)] \\
&= \prod_{j=0}^{3} [\nu_2(e, g_{3j}, g_{3(j+1)})]
\end{aligned}
$$

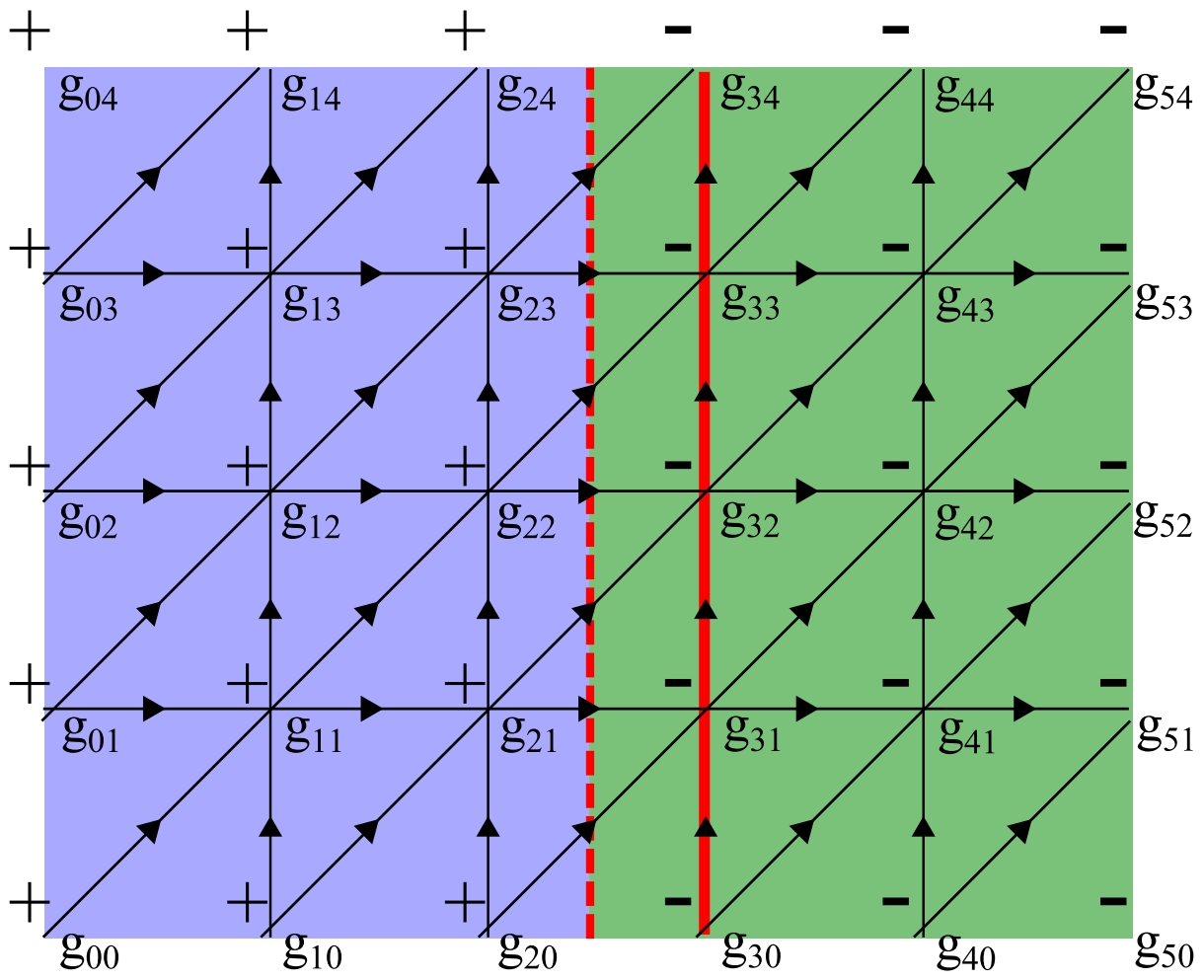

Fig. C.1. (Color online.) The wavefunction for the 2D $G \times Z_2^T$-symmetric SPT (constructed from Eq. (C.11)) with frozen configuration of the $Z_2^T$ variable (denoted by '+' (blue) and '−' (green) on each site). Upon examining the dependence of such wavefunction on the unfrozen $g_i \in G$ on each site it is noted that the value is the same as the wavefunction of a 1D $G$-symmetric SPT living on the solid red line, which is the domain wall (dashed red line) slightly displaced. Here the top and bottom edges are identified by the periodic boundary condition.

Thus the SPT wavefunctions constructed using cocycles satisfying equation (C.1) are indeed decorated domain wall wavefunctions.

## Appendix D. A $G \times Z_2^T$ invariant boundary subspace of the $(d+1)$-dimensional $G \times Z_2^T$ symmetric SPT that is transplantable to $d$ dimension

According to Eq. (B.1) and Eq. (C.1)

$$
\begin{aligned}
|\rho_\mu g_\mu\rangle_B &= \sum_{\rho_0, g_0} \prod_{\Delta} [\nu_{d+2}(e, \rho_0 g_0, \{\rho_\mu g_\mu\}_\Delta)]^{\sigma(\Delta)} |\rho_0 g_0, \{\rho_\mu g_\mu\}\rangle \\
&= \sum_{\rho_0, g_0} \prod_{\Delta} \left\{ [\nu_{d+1}(g_0, \{g_\mu\}_\Delta)]^{(\rho_0-1)/2} \right\}^{\sigma(\Delta)} |\rho_0 g_0, \{\rho_\mu g_\mu\}\rangle \\
&:= \sum_{\rho_0, g_0} \chi(\rho_0, g_0, \{g_\mu\}) |\rho_0 g_0, \{\rho_\mu g_\mu\}\rangle
\end{aligned}
\tag{D.1}
$$

It is important to note

$$
\chi(\rho_0, g_0, \{g_\mu\}) = \prod_{\Delta} \left\{ [\nu_{d+1}(g_0, \{g_\mu\}_\Delta)]^{(\rho_0-1)/2} \right\}^{\sigma(\Delta)}
$$

is independent of $\{\rho_\mu\}$. Therefore

$$|\{g_\mu\}\rangle_B := \frac{1}{(2|G|)^{1/2}} \frac{1}{2^{N/2}} \sum_{\{\rho_\mu\}} |\{\rho_\mu g_\mu\}\rangle_B, \tag{D.2}$$

where $N$ is the number of boundary sites, form an orthonormal basis

$$_B\langle\{g'_\mu\}|\{g_\mu\}\rangle_B = \prod_\mu \delta_{g'_\mu, g_\mu} \tag{D.3}$$

for the sub-Hilbert space spanned by

$$|\rho_0\rangle|g_0\rangle \prod_\mu \left( \frac{|\rho_\mu = +1\rangle + |\rho_\mu = -1\rangle}{\sqrt{2}} \right) |g_\mu\rangle.$$

The subspace spanned by $\{|\{g_\mu\}\rangle_B\}$ is isomorphic to that spanned by the usual site basis $\{|\{g_\mu\}\rangle\}$ for $G$-symmetric SPTs in one dimension lower ($d$ dimensions).

Since

$$\nu_{d+2}(g, e, g\rho_2 g_2, \ldots, g\rho_{d+2} g_{d+2}) = \nu_{d+1}(e, gg_2, \ldots, gg_{d+2})^0 = 1 \tag{D.4}$$

for $g \in G$, Eq. (B.6) implies

$$S_g|\{g_\mu\}\rangle_B = |\{gg_\mu\}\rangle_B, \tag{D.5}$$

i.e., the boundary basis $\{|\{g_\mu\}\rangle_B\}$ transform in exactly the same way as the usual site basis under group $G$. However, the $Z_2^T := \{+1, -1\}$ group transforms the boundary basis differently:

$$\begin{aligned} &S_{+1}|\{g_\mu\}\rangle_B = |\{g_\mu\}\rangle_B \\ &S_{-1}|\{g_\mu\}\rangle_B = \phi(\{g_\mu\}) K |\{g_\mu\}\rangle_B, \quad \text{where} \\ &K = \text{complex conjugation and} \\ &\phi(\{g_\mu\}) = \prod_\Delta [\nu_{d+1}(e, \{g_\mu\}_\Delta)]^{\sigma(\Delta)}. \end{aligned} \tag{D.6}$$

Because $\nu_{d+1}$ is a pure phase

$$|\phi(\{g_\mu\})|^2 = 1 \tag{D.7}$$

Eqs. (D.5) and (D.6) are the basic transformation laws of the boundary basis.

### *D.1. Breaking the $Z_2^T$ symmetry and the resulting G-symmetric SPT*

Let's come back to the basis defined by Eq. (D.2) and their transformation law, Eq. (D.6), under $G \times Z_2^T$. Now consider the following $G$-symmetric boundary Hamiltonian

$$H_0 = -J \sum_\mu \sum_{g_\mu, g'_\mu} |\{g'_\mu\}\rangle_{B\,B}\langle\{g_\mu\}|, \tag{D.8}$$

where $J > 0$ (and can be taken to very large values). Under the action of $S_g$

$$S_g H_0 S_g^{-1} = H_0 \tag{D.9}$$

while under the action of $Z_2^T$ transformation it becomes

$$S_{-1}H_0S_{-1}^{-1} = -J\sum_{\mu}\sum_{g_\mu,g'_\mu}\frac{\phi(\{g'_\mu\})}{\phi(\{g_\mu\})}\overline{|\{g'_\mu\}\rangle_B}\ \overline{_B\langle\{g_\mu\}|} := H_1 \tag{D.10}$$

where $\overline{|\ \rangle_B}$ stands for the complex conjugate. Eq. (D.10) has exactly the form of (A.4), namely the Hamiltonian for the $d$-dimensional non-trivial (assuming $\nu_{d+1}$ is non-trivial and cannot be expressed as the square of another $d+1$ cocycle) $G$-symmetric SPT. We note that $H_1$ is also invariant under the action of $S_g$, i.e.,

$$S_gH_1S_g^{-1} = H_1. \tag{D.11}$$

Moreover due to Eq. (D.7)

$$S_{-1}H_1S_{-1}^{-1} = H_0. \tag{D.12}$$

The Hamiltonian $(H_0 + H_1)/2$ is symmetric under $G \times Z_2^T$, then based on the theorem of Appendix E we conclude that either its spectrum is gapless or the $G \times Z_2^T$ symmetry is spontaneously broken. In Appendix F we give an example where $(H_0 + H_1)/2$ is gapless.

## Appendix E. A Lieb–Schultz–Mattis type theorem

In this section we present a proof stating that a $d$-dimensional system with $G \times Z_2^T$ symmetry (where $Z_2^T$ acts according to Eq. (D.6)) cannot be gapped without degeneracy.

**Proposition.** *Let $|\{g_i\}\rangle$ ($g_i \in G$) be the site basis of a $d$-dimensional lattice problem, and under the group $G$ they transform as*

$$S_g|\{g_i\}\rangle = |\{gg_i\}\rangle. \tag{E.1}$$

*Let there be an additional group $Z_2^T = \{+1, -1\}$ which acts on the basis as*

$$S_{+1}|\{g_i\}\rangle = |\{g_i\}\rangle$$
$$S_{-1}|\{g_i\}\rangle = \phi(\{g_i\})K|\{g_i\}\rangle. \tag{E.2}$$

*In addition we assume $\phi(\{g_i\})$ is the ground state wavefunction of a $d$-dimensional SPT constructed out of cocycle $\nu_{d+1}$ which cannot be expressed as the square of another cocycle, i.e.,*

$$\phi(\{g_i\}) = \prod_{\triangle}\nu_{d+1}(e, \{g_i\}_\triangle)^{\sigma(\triangle)}. \tag{E.3}$$

*Then it follows that it is impossible to find a local $G \times Z_2^T$-symmetric Hamiltonian which possesses an unique gapped ground state without breaking any symmetry.*

The situation described above arises naturally at the boundary of a $(d + 1)$-dimensional $G \times Z_2^T$ symmetric SPT. It can also be regarded as a $d$-dimensional problem with a $G$ symmetry as well as a non-local $Z_2^T$ symmetry. In either case the $Z_2^T$ symmetry ensures the impossibility to have a totally symmetric gapped ground state without breaking any symmetry. This theorem is similar to the Lieb–Schultz–Mattis theorem [26–28] for translationally invariant spin-1/2 chain. The difference is the group $G \times Z_2^T$ can be discrete.

We prove the above proposition by reductio ad absurdum. Let's assume it is possible to find an unique $G \times Z_2^T$ symmetric ground state that is separated from all excited states by an energy gap. Let $|\psi\rangle$ be such a ground state:

$$|\psi\rangle = \sum_{\{g_i\}} \chi(\{g_i\})|\{g_i\}\rangle. \tag{E.4}$$

Since $|\psi\rangle$ is symmetric under $G$ it must lie in certain equivalent class of a $G$-symmetric SPT having the fixed point form for $\chi$

$$\chi(\{g_i\}) = \prod_{\triangle} \tilde{\nu}_{d+1}(e, \{g_i\}_\triangle)^{\sigma(\triangle)}. \tag{E.5}$$

Since $|\psi\rangle$ is also invariant under $S_{-1}$

$$S_{-1}|\psi\rangle = \sum_{\{g_i\}} \chi^*(\{g_i\})\phi(\{g_i\})|\{g_i\}\rangle = |\psi\rangle = \sum_{\{g_i\}} \chi(\{g_i\})|\{g_i\}\rangle. \tag{E.6}$$

Since $\{g_i\}$ is an orthonormal set we must have

$$\chi^*(\{g_i\})\phi(\{g_i\}) = \chi(\{g_i\}), \tag{E.7}$$

or

$$\phi(\{g_i\}) = \chi(\{g_i\})^2 = \prod_{\triangle} \left[\tilde{\nu}_{d+1}(e, \{g_i\}_\triangle)^2\right]^{\sigma(\triangle)}. \tag{E.8}$$

This contradicts the assumption that $\phi$ in Eq. (E.3) cannot be constructed from the square of another cocycle. Therefore this $d$-dimensional problem must be either gapless or it spontaneously break the $G \times Z_2^T$ symmetry.

## Appendix F. Gapless $Z_2 \times Z_2 \times Z_2^T$ symmetric Hamiltonian in 1D and the transition between the $Z_2 \times Z_2$ SPTs

The $Z_2 \times Z_2$-symmetric SPTs in 1D is classified by $H^2(Z_2 \times Z_2, U(1)) = Z_2$. Let $g = (\sigma, \tau) \in Z_2 \times Z_2$ where $\sigma = \pm 1$, $\tau = \pm 1$. Following Appendix A, the bulk Hamiltonian is

$$H_1 = -J \sum_i B_i, \tag{F.1}$$

where

$$\langle\{g'_k\}|B_i|\{g_k\}\rangle = \frac{\nu_2(g_{i-1}, g_i, g'_i)}{\nu_2(g_i, g'_i, g_{i+1})} \prod_{k\neq i} \delta_{g'_k, g_k}. \tag{F.2}$$

The trivial cocycle is $\nu_2 \equiv 1$, hence the Hamiltonian associated with the trivial SPT is

$$H_0 = -J \sum_i (\sigma_i^x + \tau_i^x). \tag{F.3}$$

The nontrivial cocycle is

$$\nu_2(e, \sigma_1\tau_1, \sigma_2\tau_2) = \tau_1^{(1-\sigma_1\sigma_2)/2}, \tag{F.4}$$

hence

$$\frac{\nu_2(g_{i-1}, g_i, g_i')}{\nu_2(g_i, g_i', g_{i+1})} = (\tau_{i-1}\tau_i)^{(\frac{1-\sigma_i\sigma_i'}{2})}(\tau_i\tau_i')^{(\frac{1-\sigma_i'\sigma_{i+1}}{2})}$$
$$= \begin{cases} 1 & \text{if } (\sigma_i', \tau_i') = (\sigma_i, \tau_i) \\ \sigma_i\sigma_{i+1} & \text{if } (\sigma_i', \tau_i') = (\sigma_i, -\tau_i) \\ \tau_{i-1}\tau_i & \text{if } (\sigma_i', \tau_i') = (-\sigma_i, \tau_i) \\ \sigma_i\sigma_{i+1}\tau_{i-1}\tau_i & \text{if } (\sigma_i', \tau_i') = (-\sigma_i, -\tau_i) \end{cases} \tag{F.5}$$

Therefore the Hamiltonian associated with the non-trivial SPT is

$$H_1 = -J\sum_i \Big[1 + \tau_{i-1}^z\sigma_i^x\tau_i^z + \sigma_i^z\tau_i^x\sigma_{i+1}^z + (\tau_{i-1}^z\sigma_i^x\tau_i^z)(\sigma_i^z\tau_i^x\sigma_{i+1}^z)\Big]. \tag{F.6}$$

Note that each term of the Hamiltonian commutes with all others. In fact because the fourth term is the product of the 2nd and 3rd terms we may drop the constant term and the fourth term without changing the ground state wavefunction or closing the gap. Thus $H_0$ and the simplified $H_1$ read

$$H_0 = -J\sum_i(\sigma_i^x + \tau_i^x)$$
$$H_1 = -J\sum_i(\tau_{i-1}^z\sigma_i^x\tau_i^z + \sigma_i^z\tau_i^x\sigma_{i+1}^z). \tag{F.7}$$

Now enlarge the symmetry group to $Z_2 \times Z_2 \times Z_2^T$ and go through the steps in Appendix C it is straightforward to show is given by

$$S_{-1} = \prod_i \left(\tau_i^z\right)^{(1-\sigma_i^z\sigma_{i+1}^z)/2} K, \tag{F.8}$$

which transforms $H_0$ and $H_1$ into each another.

The critical Hamiltonian $(H_0 + H_1)/2$ can be solved exactly by the Jordan–Wigner transformation

$$\sigma_i^x = 2n_{2i-1} - 1, \quad \tau_i^x = 2n_{2i} - 1, \quad n_i = f_i^\dagger f_i$$
$$\sigma_i^z = (f_{2i-1}^\dagger + f_{2i-1})e^{i\pi\sum_{j<2i-1} n_j}$$
$$\tau_i^z = (f_{2i}^\dagger + f_{2i})e^{i\pi\sum_{j<2i} n_j}, \tag{F.9}$$

where $f_i$ are fermionic operators. The Hamiltonian is further simplified by introducing the Majorana fermion operators

$$c_{2j-1} = f_j^\dagger + f_j, \quad c_{2j} = (f_j - f_j^\dagger)/i. \tag{F.10}$$

In terms of the Majorana fermion operators

$$H_0 = -J\sum_j ic_{2j-1}c_{2j}$$
$$H_1 = -J\sum_j ic_{2j-2}c_{2j+1}$$
$$H_{\text{critical}} = -\frac{J}{2}\sum_j \left[ic_{2i-1}c_{2i} + ic_{2i-2}c_{2i+1}\right]. \tag{F.11}$$

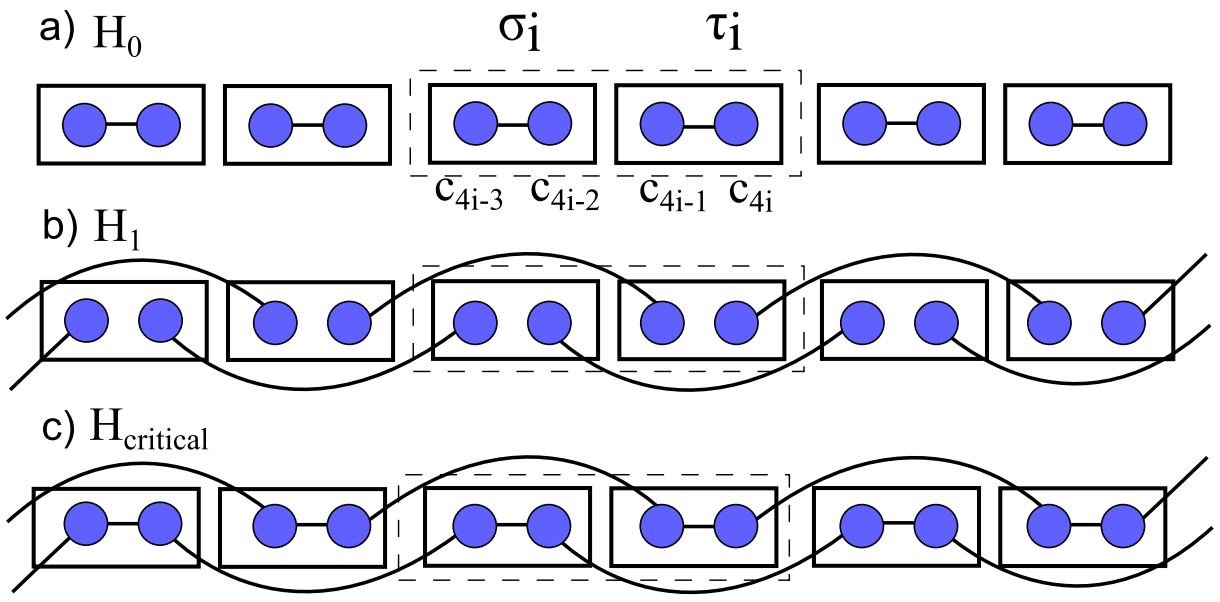

Fig. F.1. (Color online.) $H_0$ (a) and $H_1$ (b) in terms of Majorana fermions. Each bond represents a Majorana fermion hopping term. For panel (b) there are two uncoupled Majorana fermions on each of the right and left end, leading to a $2^2 = 4$ fold degeneracy. (c) A graphical representation of $H_{\text{critical}} = \frac{1}{2}(H_0 + H_1)$. There are two independent Majorana chains each contributing $1/2$ to the total central charge. The dashed lines enclose one unit cell. Each solid rectangle encloses a spin $1/2$. The blue dots denote Majorana fermions.

The hopping patterns of these Hamiltonian are shown in Fig. F.1. Because the critical Hamiltonian describes two translational-invariant gapless Majorana chain each characterized by central charge $1/2$, the total central charge of the critical Hamiltonian is 1.

Each site of the above problem can also be viewed as composing of two spin-$1/2$s each carrying a projective representation of $Z_2 \times Z_2$, or a linear representation of the quaternion group $Q_8$. The unitary transformation between the $(\sigma, \tau)$ basis discussed above and the spin-$1/2$ basis is $U = \prod_i \frac{1+i\tau_i^y}{\sqrt{2}}(\frac{1+\sigma_i^z}{2} - \frac{1-\sigma_i^z}{2}\tau_i^x)$. Under the new basis

$$
\begin{aligned}
H_0 &= \sum_i J(\sigma_i^x \tau_i^x + \sigma_i^z \tau_i^z) \\
H_1 &= \sum_i J(\tau_{i-1}^x \sigma_i^x + \tau_i^z \sigma_{i+1}^z) \\
S_\rho &= \prod_i (\frac{1+\sigma_i^z \sigma_{i+1}^z}{2} + \frac{1-\sigma_i^z \sigma_{i+1}^z}{2}\tau_i^x)K
\end{aligned}
\tag{F.12}
$$

Upon renaming $\sigma_i \to \sigma_{2i-1}$, $\tau_i \to \sigma_{2i}$ and setting $J = 1$, we obtain (8), (9) of the main text.

## Appendix G. Some details of the density matrix renormalization group calculations

We determine the ground state phase diagram and properties of the model Hamiltonian Eq. (13) by extensive and highly accurate density-matrix renormalization group (DMRG) simulations. For the DMRG simulation, we consider both a periodic boundary condition (PBC) and open boundary condition (OBC). We study many system sizes for a more reliable extrapolation to the thermodynamic limit. We keep up to $m = 1024$ states in the DMRG block with around 10 sweeps to get converged results. The truncation error is of the order $10^{-8}$ or smaller.

For the a critical theory in one dimension, the central charge of the conformal field theory can easily be extracted by fitting the entanglement entropy to the analytical form [29]

$$S(x) = \frac{c}{3\eta}\ln(x) + O(1), \tag{G.1}$$

where $x = \frac{\eta N}{\pi}\sin(\frac{\pi l}{N})$ is the so-called chord distance for a cut dividing the chain into segments of length $l$ and $N - l$, and periodic (open) boundary conditions are indicated by the parameter $\eta = 1$ or 2, respectively. Performing such a fit to the data in Fig. 9 with different system sizes, we get the central charge $c = 1$.

## Appendix H. $Z_2$ SPT in 2D

In this subsection we follow the framework set in previous sections to construct a lattice model describing phase transition between by 2D $Z_2$-symmetric SPTs. Because $H^3(Z_2, U(1)) = Z_2$, according to the cohomology group classification there are two inequivalent $Z_2$-symmetric SPTs in 2D. The non-trivial cocycle is

$$\nu_3(\sigma_1, \sigma_2, \sigma_3, \sigma_4) = (-1)^{(\frac{1-\sigma_1\sigma_2}{2})(\frac{1-\sigma_2\sigma_3}{2})(\frac{1-\sigma_3\sigma_4}{2})} \tag{H.1}$$

To construct the lattice model we consider a triangular lattice. For each site $i$ there is an Ising variable $\sigma_i := \sigma_i^z = \pm 1$. The trivial SPT Hamiltonian is

$$H_0 = -J \sum_i \sigma_i^x. \tag{H.2}$$

The non-trivial SPT Hamiltonian is

$$H_1 = -J \sum_i B_i, \tag{H.3}$$

where

$$\begin{aligned} \langle \{\sigma_i'\} | B_i | \{\sigma_i\} \rangle &= \prod_{k \neq i} \delta_{\sigma_k \sigma_k'} \times \frac{\nu_3(\sigma_3, \sigma_4, \sigma_i, \sigma_i')\nu_3(\sigma_4, \sigma_i, \sigma_i', \sigma_5)\nu_3(\sigma_i, \sigma_i', \sigma_5, \sigma_6)}{\nu_3(\sigma_3, \sigma_2, \sigma_i, \sigma_i')\nu_3(\sigma_2, \sigma_i, \sigma_i', \sigma_1)\nu_3(\sigma_i, \sigma_i', \sigma_1, \sigma_6)} \\ &= \prod_{k \neq i} \delta_{\sigma_k \sigma_k'} \times \begin{cases} 1 & \text{if } \sigma_i' = \sigma_i \\ -\left[ \Pi_{<j,k>} i^{\left(\frac{1-\sigma_j\sigma_k}{2}\right)} \right] \left[ i^{(\sum_{j=1}^{6} \sigma_j)} \right] & \text{if } \sigma_i' = -\sigma_i \end{cases} \end{aligned} \tag{H.4}$$

Here $\sigma_1, \dots, \sigma_6$ designate the Ising variables on the six neighbors of $i$ as depicted in Fig. A.1(b), and the product $\Pi_{<j,k>}$ is performed over the six links connecting site $i$ and its six nearest neighbors. After dropping a constant term we obtain

$$H_1 = \sum_i \left[ \Pi_{\langle j,k \rangle} i^{\left(\frac{1-\sigma_j\sigma_k}{2}\right)} \right] \left[ i^{(\sum_{j=1}^{6} \sigma_j)} \right] \sigma_i^x. \tag{H.5}$$

It was shown in Appendix D of Ref. [24] that $H_1$ is related to the Levin–Gu [7] Hamiltonian $H_{LG} = \sum_i \left[ \Pi_{\langle j,k \rangle} i^{\left(\frac{1-\sigma_j\sigma_k}{2}\right)} \right] \sigma_i^x$ by a local unitary transformation. The non-trivial element of the $Z_2^T$ transformation is given by

$$S_{-1} = \prod_{\Delta} (-1)^{(\frac{1-\sigma_1}{2})(\frac{1+\sigma_2}{2})(\frac{1-\sigma_3}{2})} K \tag{H.6}$$

where $\sigma_1, \sigma_2, \sigma_3$ are the ordered vertices on each triangle $\Delta$. Again $S_{-1} H_0 S_{-1}^{-1} = H_1$ and $S_{-1} H_1 S_{-1}^{-1} = H_0$.

We construct the Hamiltonian to study the phase transition in exactly the same way as in Eq. (4). $H(\lambda)$ is only solvable for $\lambda = 0$ and 1. For intermediate value of $\lambda$ it was suggested [8] numerically that there is a first-order transition at $\lambda = 1/2$.

## References

[1] X. Chen, Z.-C. Gu, Z.-X. Liu, X.-G. Wen, Symmetry protected topological orders and the group cohomology of their symmetry group, Phys. Rev. B 87 (2011) 155114.
[2] X. Chen, F. Wang, Y.-M. Lu, D.-H. Lee, Critical theories of phase transition between symmetry protected topological states and their relation to the gapless boundary theories, Nucl. Phys. B 873 (2013) 248.
[3] A. Kapustin, Symmetry protected topological phases, anomalies, and cobordisms: beyond group cohomology, arXiv:1403.1467, 2014.
[4] Y.-M. Lu, D.-H. Lee, Quantum phase transitions between bosonic symmetry-protected topological phases in two dimensions: emergent QED 3 and anyon superfluid, Phys. Rev. B 89 (19) (2014) 195143.
[5] T. Grover, A. Vishwanath, Quantum phase transition between integer quantum hall states of bosons, Phys. Rev. B 87 (4) (2013) 045129.
[6] X. Chen, Y.-M. Lu, A. Vishwanath, Symmetry-protected topological phases from decorated domain walls, Nat. Commun. 5 (2014).
[7] M. Levin, Z.-C. Gu, Braiding statistics approach to symmetry-protected topological phases, Phys. Rev. B 86 (2012) 115109.
[8] S.C. Morampudi, C. Von Keyserlingk, F. Pollmann, Numerical study of a transition between Z2 topologically ordered phases, Phys. Rev. B 90 (3) (2014) 035117.
[9] X.-G. Wen, Topological orders and edge excitations in fractional quantum hall states, Adv. Phys. 44 (5) (1995) 405–473.
[10] C. Wang, A.C. Potter, T. Senthil, Gapped symmetry preserving surface state for the electron topological insulator, Phys. Rev. B 88 (11) (2013) 115137.
[11] C. Wang, T. Senthil, Interacting fermionic topological insulators/superconductors in three dimensions, Phys. Rev. B 89 (19) (2014) 195124.
[12] F. Burnell, X. Chen, L. Fidkowski, A. Vishwanath, Exactly soluble model of a 3d symmetry protected topological phase of bosons with surface topological order, arXiv:1302.7072, 2013.
[13] M.A. Metlitski, C. Kane, M. Fisher, A symmetry-respecting topologically-ordered surface phase of 3d electron topological insulators, arXiv:1306.3286, 2013.
[14] X. Chen, L. Fidkowski, A. Vishwanath, Symmetry enforced non-abelian topological order at the surface of a topological insulator, Phys. Rev. B 89 (16) (2014) 165132.
[15] P. Bonderson, C. Nayak, X.-L. Qi, A time-reversal invariant topological phase at the surface of a 3d topological insulator, J. Stat. Mech. Theory Exp. 2013 (09) (2013) P09016.
[16] A. Vishwanath, T. Senthil, Physics of three-dimensional bosonic topological insulators: surface-deconfined criticality and quantized magnetoelectric effect, Phys. Rev. X 3 (1) (2013) 011016.
[17] L. Fidkowski, X. Chen, A. Vishwanath, Non-abelian topological order on the surface of a 3d topological superconductor from an exactly solved model, Phys. Rev. X 3 (4) (2013) 041016.
[18] S.R. White, Density matrix formulation for quantum renormalization groups, Phys. Rev. Lett. 69 (19) (1992) 2863.
[19] F. Haldane, Nonlinear field theory of large-spin Heisenberg antiferromagnets: semiclassically quantized solitons of the one-dimensional easy-axis Néel state, Phys. Rev. Lett. 50 (15) (1983) 1153.
[20] I. Affleck, T. Kennedy, E.H. Lieb, H. Tasaki, Rigorous results on valence-bond ground states in antiferromagnets, Phys. Rev. Lett. 59 (7) (1987) 799.
[21] I. Affleck, F. Haldane, Critical theory of quantum spin chains, Phys. Rev. B 36 (10) (1987) 5291.
[22] Z. Bi, A. Rasmussen, C. Xu, Classification and description of bosonic symmetry protected topological phases with semiclassical nonlinear sigma models, arXiv:1309.0515, 2013.
[23] J. Oon, G.Y. Cho, C. Xu, Two-dimensional symmetry-protected topological phases with PSU($N$) and time-reversal symmetry, Phys. Rev. B 88 (1) (2013) 014425.
[24] A. Mesaros, Y. Ran, A classification of symmetry enriched topological phases with exactly solvable models, Phys. Rev. B 87 (2012) 155115.
[25] X. Chen, Z.-X. Liu, X.-G. Wen, Two-dimensional symmetry-protected topological orders and their protected gapless edge excitations, Phys. Rev. B 84 (23) (2011) 235141.

[26] E. Lieb, T. Schultz, D. Mattis, Two soluble models of an antiferromagnetic chain, Ann. Phys. 16 (3) (1961) 407–466.
[27] I. Affleck, E. Lieb, A proof of part of Haldane's conjecture on spin chains, Lett. Math. Phys. 12 (1) (1986) 57–69.
[28] C. Xu, A.W. Ludwig, Nonperturbative effects of a topological theta term on principal chiral nonlinear sigma models in $2 + 1$ dimensions, Phys. Rev. Lett. 110 (20) (2013) 200405.
[29] P. Calabrese, J. Cardy, Entanglement entropy and quantum field theory, J. Stat. Mech. Theory Exp. 2004 (06) (2004) P06002.